\documentclass[11pt,a4paper]{article}
\usepackage[utf8]{inputenc}
\usepackage[T1]{fontenc}
\usepackage{geometry}
\usepackage{array}
\usepackage{multicol}
\usepackage{comment}
\usepackage[thinc]{esdiff}
\usepackage{amsopn,amsmath,amssymb,amsthm,amstext,amsfonts,algorithmic,psfrag}
\usepackage{amssymb}
\usepackage{enumitem}
\usepackage{mathtools}
\usepackage[english]{babel}
\usepackage{hyperref}
\hypersetup{colorlinks=true, citecolor=blue, linkcolor=blue}
\usepackage[round]{natbib}
\setcitestyle{aysep={,}} 
\usepackage{authblk}
\usepackage{graphicx}
\usepackage{mwe}    
\usepackage{subfigure} 
\usepackage{float}
\usepackage{caption}
\usepackage{algorithm}
\usepackage{algorithmic,latexsym}
\usepackage{placeins,enumitem}
\renewcommand{\thetable}{\arabic{table}}

\usepackage{eso-pic}
\newcommand\AtPageUpperMyright[1]{\AtPageUpperLeft{%
 \put(\LenToUnit{0.5\paperwidth},\LenToUnit{-1cm}){%
     \parbox{0.5\textwidth}{\raggedleft\fontsize{9}{11}\selectfont #1}}%
 }}%
\newcommand{\conf}[1]{%
\AddToShipoutPictureBG*{%
\AtPageUpperMyright{#1}
}
}

\title{\LARGE \bf
Critical Medical Resource Allocation during COVID-19 Pandemic
}

\author[1]{Shuvrangshu Jana}
\author[2]{Rudrashis Majumder}
\author[3]{Debasish Ghose}

\affil[1]{Post-doctoral Fellow, Department of Aerospace Engineering, Indian Institute of Science, Bangalore,
		{\tt\small shuvrangshuj@iisc.ac.in}}

\affil[2]{Ph.D. Student, Department of Aerospace Engineering, Indian Institute of Science, Bangalore, 
		{\tt\small rudrashism@iisc.ac.in}}

\affil[3]{Professor, Department of Aerospace Engineering, Indian Institute of Science, Bangalore, 
		{\tt\small dghose@iisc.ac.in}}		
		
\date{}                     
\begin{document}

\maketitle
\conf{$5^{th}$ World Congress on  Disaster Management\\ IIT Delhi, New Delhi, India, 24-27 November 2021}



\begin{abstract}
 In this paper, an optimal resource allocation framework is proposed for the allocation of critical medical resources among different units during a pandemic. The framework is developed by considering the dynamics of Pandemic, hierarchical government structure, and non-uniformity of unit resource requirement among different units.  
 The cost function is designed to minimize the difference between the demand, actual allocation, and ideal allocation, where ideal allocation for a region is considered based on the predicted active cases in a fraction of predicted total active cases of all regions. Different cost functions are used at a different level of organization based on the available information. The model can also accommodate severity of disaster in a region in this framework. A sample allocation case study is presented for the allocation of oxygen for different states of India. 
\end{abstract}


\section{Introduction}

The sudden outbreak of the COVID-19 pandemic in late 2019 has stopped society from functioning normally because of losses of life and economies worldwide. The mitigation of COVID-19 needs a series of logistic support and medical infrastructure. In such emergency scenarios, the importance of an automated decision support system (DSS) can not be ignored. One of the major response activities needed from the COVID-19 management authority is the distribution of medical resources like testing kits, ventilators, oxygen cylinders, etc. This is a vital step to manage a pandemic, especially at early stages when the resources are highly scarce. Resource allocation of critical items should be adaptive to the dynamics of the disaster and the evolving ground scenarios. However, manual decision-making may involve inaccuracies and bias, which can make the situation worse. Hence, the automated DSS can play a significant role in the medical resource allocation during the COVID-19 pandemic. Several papers like \citep{jenvald2007simulation, araz2013simulation, araz2013integrating, fogli2013knowledge, shearer2020infectious} mainly focus on designing DSS in order to build preparedness for a pandemic and necessary response operations. The complexities associated with the dynamic decision-making for an influenza pandemic are addressed in \citep{arora2012decision}. The article \citep{phillips2020supporting} uses analytics to explore decision-making in the stressful condition imposed by the COVID-19 pandemic. A DSS is proposed in \citep{guler2020decision} to address a shift scheduling problem for the physicians during COVID-19 pandemic. This problem is solved using a mixed integer programming model. \cite{govindan2020decision} presents a problem of demand management in healthcare supply chain to mitigate COVID-19. The paper \citep{sharma2020multi} uses a multi-agent setup for COVID-19 healthcare for continuous monitoring of the patients. A few other papers like \citep{marques2021prediction, hashemkhani2020application} also deals with the decision support systems to handle COVID-19.
  
Mitigation of a disaster needs fair and efficient allocation of different critical resources. However, scarcity of resources compared to the requirement makes the job of resource allocation critical for the disaster management authorities. Research concerned to the allocation and distribution of resources involves mathematical techniques like game theory, machine learning, combinatorial optimization, heuristic methods, etc. Hierarchical multi-agent reinforcement learning is proposed for optimal resource allocation for real-time adaptation to dynamic events and evolving disaster \citep{vereshchaka2019dynamic}. Another paper that deals with hierarchical resource allocation is disaster response is \citep{jana2021decision}. A multi-objective cellular genetic algorithm is used to minimize the disaster losses and to satisfy the demand at the earliest \citep{wang2020emergency}. In \citep{celik2017stochastic}, a two stage stochastic optimization model is proposed for optimal resource distribution location and allocation for a large-scale disaster.  In \citep{sheu2007emergency},  relief allocation in a disaster scenario is presented considering time-varying relief demand using forecast model, a grouping of affected areas using fuzzy logic, relief distribution priority using urgency attribute, and optimal relief supply considering the transportation cost. The papers \citep{gupta2006social, ranganathan2007automated} formulates resource scheduling problems based on non-cooperative game with multiple disaster locations and resource stations. In \citep{majumder2019game}, non-cooperative game is used to find the equilibrium solution for a resource allocation problem during disaster.

Resource allocation literature related to  COVID-19  mostly highlighted ethical challenges in allocating scarce resources like ventilators among patients \citep{dawson2020ethics, parsons2020best, laventhal2020ethics, siraj2020infectious, worby2020face}. There is not much literature available online that studies the optimization of logistics and supply chain for the allocation of resources in the COVID-19 scenario. \citep{prakash2020minimal} mainly focuses on estimating the number of infected and deceased people and derives the medical resource requirement in a location. \cite{herold2021COVID} discusses the supply chain model and logistics management in the light of COVID-19. \citep{montero2020age} studies the impact of different parameters other than the age of the patient, which can determine the allocation of the healthcare resources.

  For optimal use of scarce resources, it is necessary to have a centralized resource allocation framework to allocate the resources among different units. In general, resources are distributed in a hierarchical manner, and the allocation factors could be different at different levels. The total number of active cases reflects the average medical requirement at a region. Detailed calculation of number and type of patients could provide the detailed requirement of resources; however, this exercise will not be possible at a higher level. In a real scenario, allocation based solely on the number of active cases in a region might be wrong as different units could have different testing strategies, reporting frameworks, and medical protocols. Therefore, a resource allocation framework should be designed considering the feasibility of desired input at that level and also accommodating the variation in requirement per active cases of different units.
 
   In this paper, a hierarchical resource allocation framework for critical medical items is designed in the Indian context, where the optimization function is designed considering the information flow at that level of organization.  Allocation architecture is developed based on the hierarchical administrative structure of the Indian government.  Resources are allocated based on dynamic factors such as the predicted value of the active cases and the severity of an area that could be accommodated using lockdown status and the growth rate of test positive ratio. Two different cost functions are proposed for resource allocation; one for the lowest level and the other for the upper levels. A case study for allocation of oxygen by the center to states in a  resource-constrained scenario is shown considering the state's actual demands and the state's reported active cases. 
   
  The rest of the paper is described as follows:  Important variables are described in Section \ref{nomenclature}.  The mathematical framework of resource allocation is developed in Section \ref{maths}. A case study for the allocation of oxygen among different states is presented in Section \ref{casestudy}. The concluding remarks are given in Section \ref{conclusion}.

\section{Nomenclature} \label{nomenclature}

\begin{itemize}
\item[] $\delta_{i}$ =  Disaster criticality of state $i$

\item[] $ D_{ij} $ =   Demand  of resource $j$ raised  by state $i$

\item[] $ A_{ij}^{0} $ =  Allocation of $j^{\text{th}}$ resource to  $i^{\text{th}}$ state by centre based on state's active case without considering demand of individual states
       
\item[] $T_{j}$ =  Total amount of $j^{th}$ resource available to centre
  
\item[] $\alpha_{ij}$= Fraction of  resource $j$ allocated to state $i$

\end{itemize}

 \section{Mathematical formulation} \label{maths}
 
Considering the resource allocation structure in India, critical items like oxygen are distributed by the center to different states, and then the state further distributes to its districts, and finally, districts allocate to different hospitals in its area. The number of total confirmed cases, daily cases, recovered cases, and deceased cases are reported by states to center. The states calculate their overall demand based on the requirement of each district, and demand of each district is calculated at the micro-level considering the COVID-beds status and consumption of medical items at ground levels.  Allocation is formulated in 3 stages: a) Center to states, b) State to districts,  c) District to hospitals. 
 
\subsection{ Centre to state allocation}
Comparison of the total number of active cases of two different states might not reflect the actual difference between the current state of pandemic in the two states, as the testing strategy, demography, discharge criteria can vary significantly. So, some states might require some resources that could not be estimated only from the active cases.  So, the weightage for allocation is considered based on the active cases; however, the demands of individual states are taken into account for the resource allocation. The demand of individual states could be adjusted before the allocation algorithm for outlier cases. Apart from active cases,  test positive ratio, lockdown status, and death rate could also be considered in the weightage. 
  
The allocation at the center to state is performed to reduce the sum of the square of the error between the actual allocation with state demand and ideal allocation by the center. Actual allocation by the center is the amount allocated by the center to states.  Ideal allocation by the center is defined as an amount without considering the state's specific demands. The following cost function is used for derivation of allocation matrix for different states. 
 
  \begin{equation}
      \text{Minimize} \quad  J= \sum_{i} \delta_{i} \bigg(\frac{D_{ij}-\alpha_{ij} T_{j}}{D_{ij}}\bigg)^2 +  \sum_{i} \delta_{i}  \bigg( \frac{A_{ij}^{0}-\alpha_{ij} T_{j}}{A_{ij}^0}           \bigg)^2   \label{opt_1}
  \end{equation}

 \noindent   subject to following constraints, 
   \begin{equation}
       \sum_{i} \alpha_{ij} =1  \label{constraint_1}
   \end{equation}

  In the first part of cost function $J $, $(D_{ij}-\alpha_{ij} T_j)^2$ represents the cost due to difference between the demand of state and the actual amount allocated by the centre.   The term $\delta_{i}$ is used to  consider the severity of each state into consideration. The term $\delta_{i}$  could be  represented by the state positive ratio of the status of lockdown. The term $D_{ij}^2$ in the denominator is used for normalization as the demands from different states  varies significantly.  In the second term, $(A_{ij}^{0}-\alpha_{ij} T_{j})^2$ represents the cost due to difference between ideal allocation by centre and the actual allocation by centre.  The  term $(A_{ij}^{0})^2$  is used for normalization. The optimal value of $\alpha_{ij}$ needs to be derived to obtain the allocation  factor of  $i^{\text{th}}$  state for $j^{\text{th}}$  resource.   The equality constraint in (\ref{constraint_1}) is used as all resources are distributed  among states.  The allocation of $j^{\text{th}}$ resource to  state $i$ is obtained as, 
   \begin{equation}
       O_{ij}= \alpha_{ij}^{\text{opt}} T_{j}
   \end{equation}
   where, $\alpha_{ij}^{\text{opt}} $ is the optimal value of $\alpha_{ij} $ after solving the Eqns.  (\ref{opt_1}) and  (\ref{constraint_1}).

\noindent  Consider the following  augmented cost function, 
 
 \begin{equation}
     \tilde{J} = \sum_{i} \delta_{i} \bigg(\frac{D_{ij}-\alpha_{ij} T_{j}}{D_{ij}}\bigg)^2 +  \sum_{i} \delta_{i}  \bigg( \frac{A_{ij}^{0}-\alpha_{ij} T_{j}}{A_{ij}^0} \bigg)^2 + \lambda (\sum_{i} \alpha_{ij} -1)
 \end{equation}
 
\noindent  For $\tilde{J}$ to be minimum, the necessary conditions are, 
 \begin{equation}
     \frac {\partial \tilde{J}}{ \partial\alpha_{ij}} =0; \quad \text{and} \quad     \frac {\partial \tilde{J}}{ \partial \lambda} =0
 \end{equation}
 
\noindent  Using the condition, $ \frac{\partial \tilde{J}}{ \partial \alpha_{ij}} =0 $,  we get,
 
 \begin{equation}
     -2 \frac{\delta_{i}}{D_{ij}^2} (D_{ij}-\alpha_{ij} T_{j})T_{j}- 2 \frac{\delta_{i}}{(A_{ij}^{0})^2} (A_{ij}^{0}-\alpha_{ij} T_{j}) T_{j} + \lambda=0 ;   \label{diff1}
 \end{equation}

 \noindent Using the condition, $\frac {\partial \tilde{J}}{ \partial \lambda} =0,$ we get,  
 \begin{equation}
     \sum_{i} \alpha_{ij} -1 =0;  \label{diff2}
 \end{equation}
 
\noindent  So, for a given resources $j$, we have  $i+1$ number of equations, and we have unknown variables $\lambda$  and $i$ number of $\alpha_{ij} $.  Equation  (\ref{diff1}) can be simplified as, 

\begin{equation}
    2 \alpha_{ij} ( \frac{\delta_{i}}{D_{ij}^2} T_{ij}^2 + \frac{\delta_{i}}{(A_{ij}^{0})^2} T_{ij}^2) - 2( \frac{\delta_{i}}{D_{ij}} T_{ij} + \frac{1}{A_{ij}^{0}} T_{ij}) + \lambda=0  \label{alphaij}
\end{equation}

\noindent Equation  (\ref{alphaij}) can be rearranged  to obtain the value of $\alpha_{ij}$ as, 

\begin{equation}
    \alpha_{ij}= \frac{(\frac{\delta_{i}}{D_{ij}}+ \frac{1}{A_{ij}^{0}})T_{ij}-0.5\lambda}{(\frac{\delta_{i}}{D_{ij}^2}+ \frac{1}{(A_{ij}^{0})^2})T_{ij}^2}
\end{equation}

Using the value of $\alpha _{ij}$ in (\ref{diff2}), we get, 
 
\begin{equation}
    \sum_{i} \left(\frac{(\frac{\delta_{i}}{D_{ij}}+ \frac{1}{A_{ij}^{0}})T_{ij}-0.5\lambda}{(\frac{\delta_{i}}{D_{ij}^2}+ \frac{1}{(A_{ij}^{0})^2})T_{ij}^2}\right) =1
\end{equation}
 
\noindent Therefore,

\begin{equation}
     \sum_{i} \left(\frac{(\frac{\delta_{i}}{D_{ij}}+ \frac{1}{A_{ij}^{0}})T_{ij}}{(\frac{\delta_{i}}{D_{ij}^2}+ \frac{1}{(A_{ij}^{0})^2})T_{ij}^2}\right)
=1 +  \sum_{i} \left(\frac{0.5\lambda}{(\frac{\delta_{i}}{D_{ij}^2}+ \frac{1}{(A_{ij}^{0})^2})T_{ij}^2}\right) 
\end{equation}

\noindent Hence, the value of $\lambda$ is obtained as, 

\begin{equation}
    \lambda=  \frac{ 2\left( \sum_{i} \frac{(\frac{\delta_{i}}{D_{ij}}+ \frac{1}{A_{ij}^{0}})T_{ij}}{(\frac{\delta_{i}}{D_{ij}^2}+ \frac{1}{(A_{ij}^{0})^2})T_{ij}^2} -1\right)} {\sum_{i} \left(\frac{1}{(\frac{\delta_{i}}{D_{ij}^2}+ \frac{1}{(A_{ij}^{0})^2})T_{ij}^2}\right) }
\end{equation}
  
 Therefore, the final value of $\alpha_{ij}$ is, 
 
\begin{equation}
    \alpha_{ij}= \frac{(\frac{\delta_{i}}{D_{ij}}+ \frac{1}{A_{ij}^{0}})T_{ij}-\frac{ \left( \sum_{i} \frac{(\frac{\delta_{i}}{D_{ij}}+ \frac{1}{A_{ij}^{0}})T_{ij}}{(\frac{\delta_{i}}{D_{ij}^2}+ \frac{1}{(A_{ij}^{0})^2})T_{ij}^2} -1\right)} {\sum_{i} \left(\frac{1}{(\frac{\delta_{i}}{D_{ij}^2}+ \frac{1}{(A_{ij}^{0})^2})T_{ij}^2}\right) }}{(\frac{\delta_{i}}{D_{ij}^2}+ \frac{1}{(A_{ij}^{0})^2})T_{ij}^2}
\end{equation}

   

    
    
    

\subsection{State to district allocation }

  Testing strategy, medical intervention, release criteria from the hospital,  reporting of information does not vary drastically from district to district for a given state. Therefore, it can be assumed that for a given state,  the active cases of each district will reflect the status of each district. So, the allocation matrix can be formulated based on the reported active cases of each district, and a  proportional allocation could be used based on the predicted active cases of a district as a fraction of the predicted value of active cases of state. However, a similar strategy as the one followed by the centre could be followed by the states in order to accommodate individual district demand in the resource allocation framework. The weight on different districts could also be designed, including population density, demography apart from the lockdown status, and test positive ratio of each district. 
  
  \subsection{District to hospital allocation}
   It is very difficult to accommodate the dynamics of disease in the allocation of medical resources to hospitals by districts, as the prediction of patients in a hospital will have huge uncertainty. However, the current demand of each hospital can be estimated with very good accuracy, and allocation at the districts level could be modified quickly based on the situation.  Therefore, hospitals' demand could be considered the actual requirement at the district level, and optimal allocation could be performed based on the demand without including a complex prediction model of expected patients in a hospital.
  
  Let us consider an allocation scenario of  $j^{\text{th}}$ resource  in a district which has $l$ hospitals.  Let $D_{lj}$ be the demand of   $j^{\text{th}}$
   resources of $l^{\text{th}}$ hospital. Amount of $j^{\text{th}}$ resource allocated to this district is  $T^{d}$. Let the criticality  factor of $l^{\text{th}}$ is $\delta_{l}$ and  $\gamma_{lj} $ is the allocation fraction of $l^{\text{th}}$ hospital for $j^{\text{th}}$ resource.    Allocation factor $\gamma_{ij}$   is obtained   using  the following cost function.

 \begin{equation}
     \text{minimize} \quad  J^{d} = \sum_{i} \delta_{l}(\frac{D_{lj}-\gamma_{lj}T^{d}}{D_{lj}}  )^2
 \end{equation}
  
 \noindent Subject to the conditions, 
 
 \begin{equation}
     \sum_{l} \gamma_{lj} =1
 \end{equation}
  
Let us consider the augmented cost function, 

 \begin{equation}
     \text{minimize} \quad  \tilde{J^{d}} = \sum_{i} \delta_{l}(\frac{D_{lj}-\gamma_{lj}T^{d}}{D_{lj}} )^2 +\lambda^{d} (\sum_{l} \gamma_{lj} -1)
 \end{equation}  
  
 For $\tilde{J^{d}}$ to be minimum,  the necessary conditions are, 
\begin{equation}
    \frac{\partial \tilde{J^{d}}}{\partial \gamma_{ij}}= 0 \quad      \frac{\partial \tilde{J^{d}}}{\partial \lambda^{d}}= 0
\end{equation}

Using the necessary conditions and further simplification, we get, 

\begin{equation}
    \gamma_{ij}=  \frac{\frac{\delta_{l}T^{d}}{D_{lj}} -0.5 \lambda^{d}} {\frac{\delta_{lj} (T^{d})^2}{D_{lj}^2}} 
\end{equation}

\begin{equation}
    \lambda= \frac{2(\sum_{l}\frac{D_{lj}}{T^{d}}-1)}{\sum_{l}\frac{D_{lj}^2}{\delta_{l}(T^{d})^2}}
\end{equation}

Let, $\gamma_{lj}^{opt}$ is the optimal value of  $\gamma_{lj}$.  Then, 
\begin{equation}
  \gamma_{lj}^{opt}= \frac{\frac{\delta_{l}T^{d}}{D_{lj}} - \frac{(\sum_{l}\frac{D_{lj}}{T^{d}}-1)}{\sum_{l}\frac{D_{lj}^2}{\delta_{l}(T^{d})^2}} 
  }{\frac{\delta_{lj} (T^{d})^2}{D_{lj}^2}}  
\end{equation}

    
    
    

 

 \section{Case study} \label{casestudy}
  We have considered an allocation scenario of oxygen by the Indian government among its different states on $20^{\text{th}}$ April. At this stage, the second wave of COVID-19 is rising, and the states are finding it difficult to meet the demand for oxygen.  The active case history of different states is calculated using the information from "https://api.COVID19india.org/" website.  As per information from the Ministry of Home Affairs of Government of India order 40-3/2020-DM-I(A) dated 22.04.2021, the demand of different states on  $20^{\text{th}}$ April is shown in Table \ref{tab:oxygendemand}. We are considering only those states which are mentioned in the  Table \ref{tab:oxygendemand}.  The overall oxygen demand is 6595 MT. We consider a resource allocation scenario where the total available oxygen is 5000 MT. We have allocated 5000 MT using our resource allocation formulation considering states' demand and active cases.  For simplicity, we have considered an equal priority of each state without considering their lockdown status and test positive ratio; that is, the value of $\delta_{i}$ is the same for all states.

  The demand of each state in fraction of total demand,  total active cases as on $20^{\text{th}}$ April, and active cases of each state in fraction of total active cases are also shown in  Table \ref{tab:oxygendemand}. It is observed that some state's demands are not correlated with their active cases.  For example, Gujarat has a demand fraction of 15.16 with an active case fraction of 3.98; whereas,  Kerala has a demand fraction of 1.35 with an active case fraction of 6.18. Clearly, proportional allocation based on the active cases or the respective demand will not be a fair allocation. This also justifies our problem formulation of accommodating the demand and the ideal allocation simultaneously.  
 
 \begin{table}[hbt!]
\caption{\label{tab:oxygendemand}  Demand of Oxygen of different states and corresponding active cases}
\centering
\begin{tabular}{lcccc}
\hline
\hline
\bf{State} & \bf{Demand (MT)} &  $w^{D}$  & Active case &  $w^{A}$   \\ \hline
         Maharashtra  &  1500  &  22.75     &  683856  &   35.62  \\
         
         Gujarat &   1000  &     15.16    &  76500 &     3.98        \\ 
         
        Karnataka &   300  &   4.55      & 159158   &    8.29   \\
        
         Madhya Pradesh & 445  &  6.75   & 78271  & 4.08    \\
         
        Delhi &  700  &  10.61  &  85571 & 4.46  \\ 
        
        Haryana    & 180   & 2.73    & 49772 &   2.59      \\
        
        Uttar Pradesh &  800 & 12.13   & 223544 &  11.64  \\
        
        Tamil Nadu  &  200 &  3.03  & 79804  & 4.16  \\
        
        Kerala  &  89 &  1.35    & 118669 &  6.18  \\
        
        Chhattisgarh   & 215  &  3.26   & 125688 & 6.55 \\
        
        Rajasthan  & 205 &  3.11   & 85571 & 4.46 \\
        
         Telangana &  360  &   5.46  & 42853 & 2.23 \\
         
         Andhra Pradesh   & 440 & 6.67 & 53889  & 2.81 \\
         
         Uttarakhand  &  103  &  1.56  & 21014 &  1.09 \\
         
         Jammu and Kashmir  & 12 & 0.18   & 13470 & 0.70  \\
         
         Goa &  11 &  0.17 & 8241 &  0.43 \\
         
         Chandigarh & 20 & 0.30   & 3959 & 0.21   \\
         
         Himachal Pradesh  &  15  &   0.23  & 10029 & 0.52 \\

\hline
\hline
\end{tabular}
\end{table}
 
\begin{figure}[hbt!]
\centering
\subfigure[ \textbf{Maharashtra}]{%
\includegraphics[width=0.32\columnwidth,keepaspectratio]{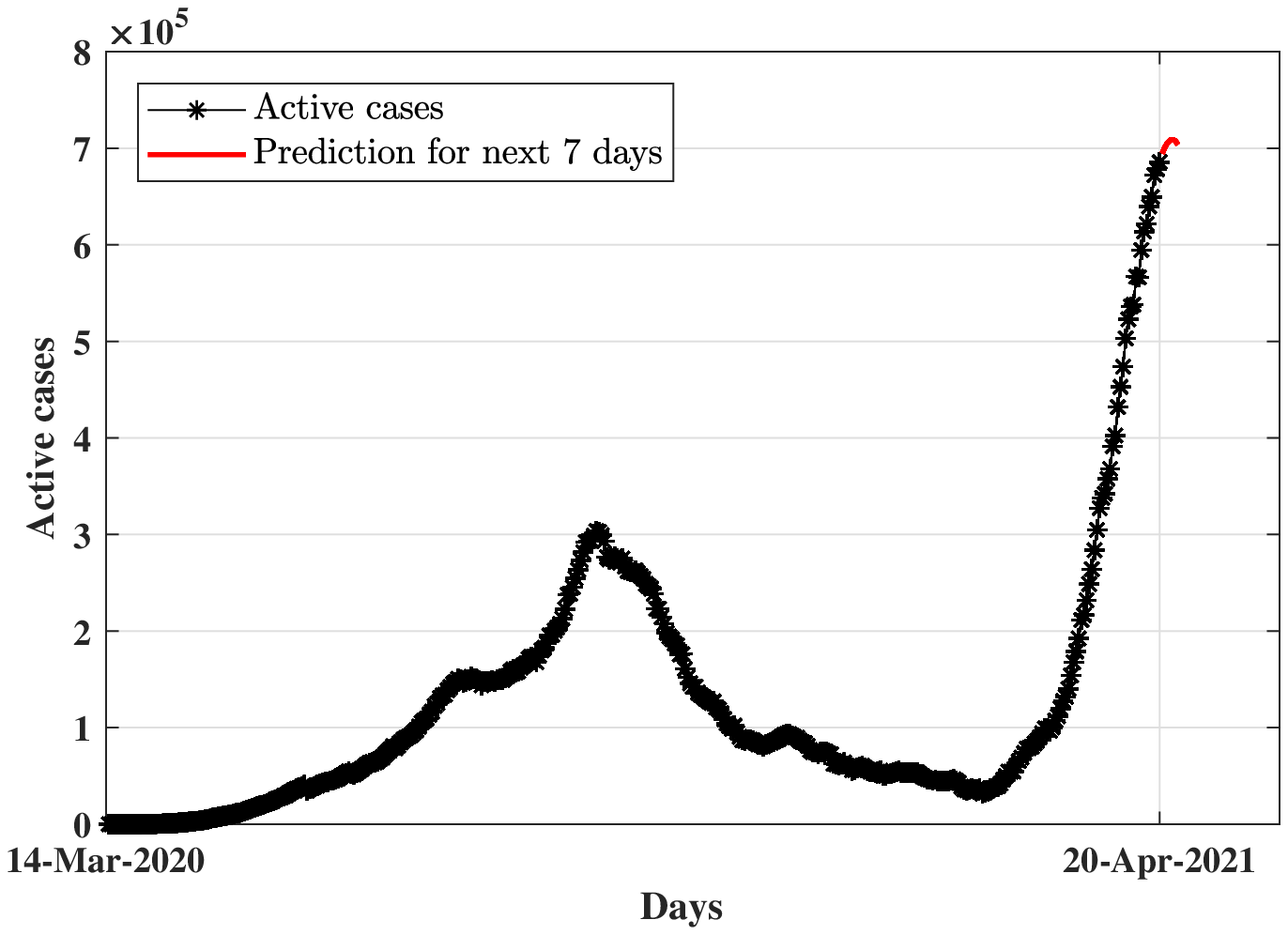}    %
\label{fig:2dtrajectorymaneuvere1}}  
\subfigure[\textbf{Gujarat }]{%
\includegraphics[width=0.32\columnwidth, height=0.25\columnwidth]{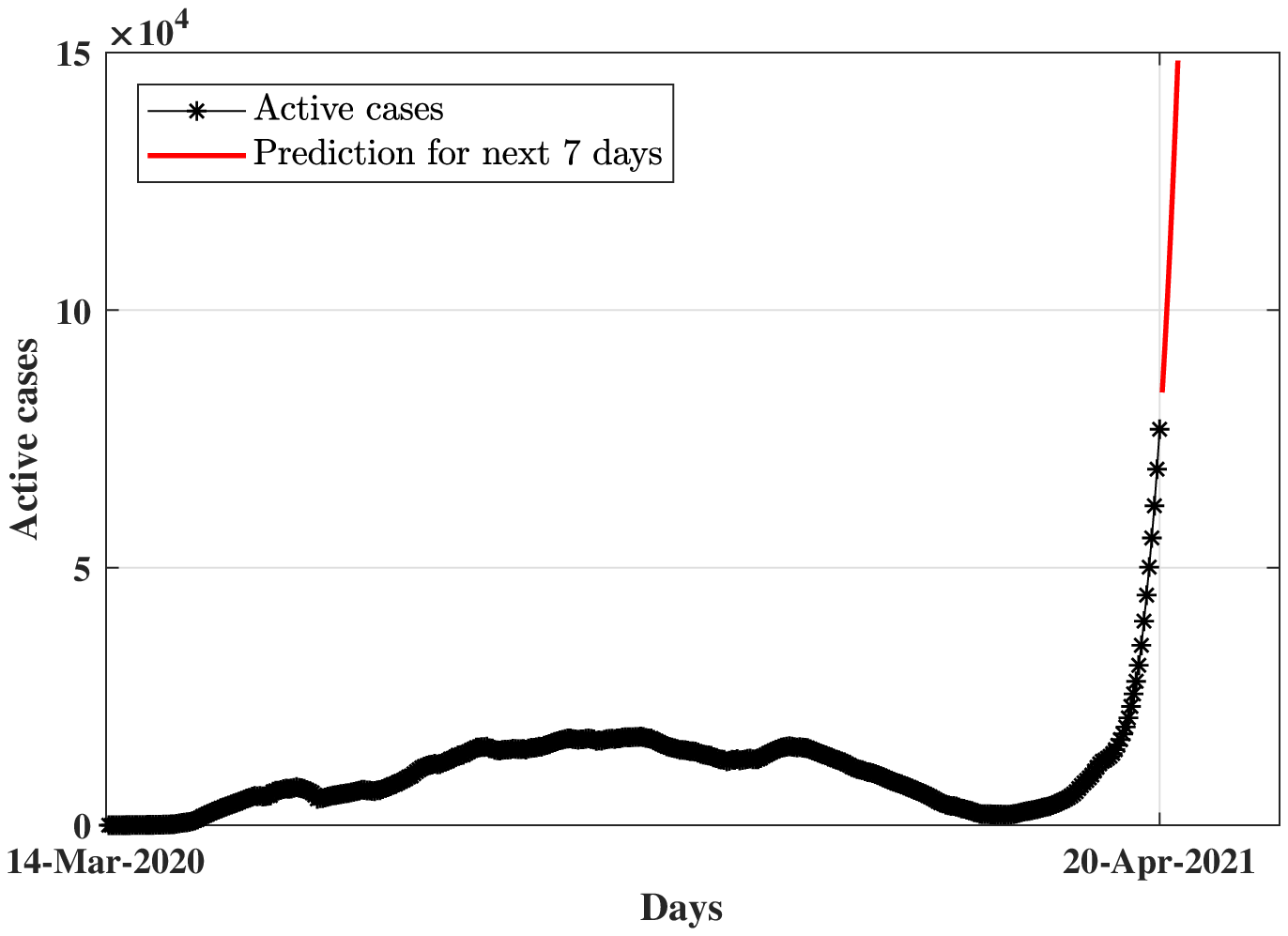}    
\label{fig:man2DR1}}
\subfigure[\textbf{ Karnataka}]{%
\includegraphics[width=0.32\columnwidth, keepaspectratio]{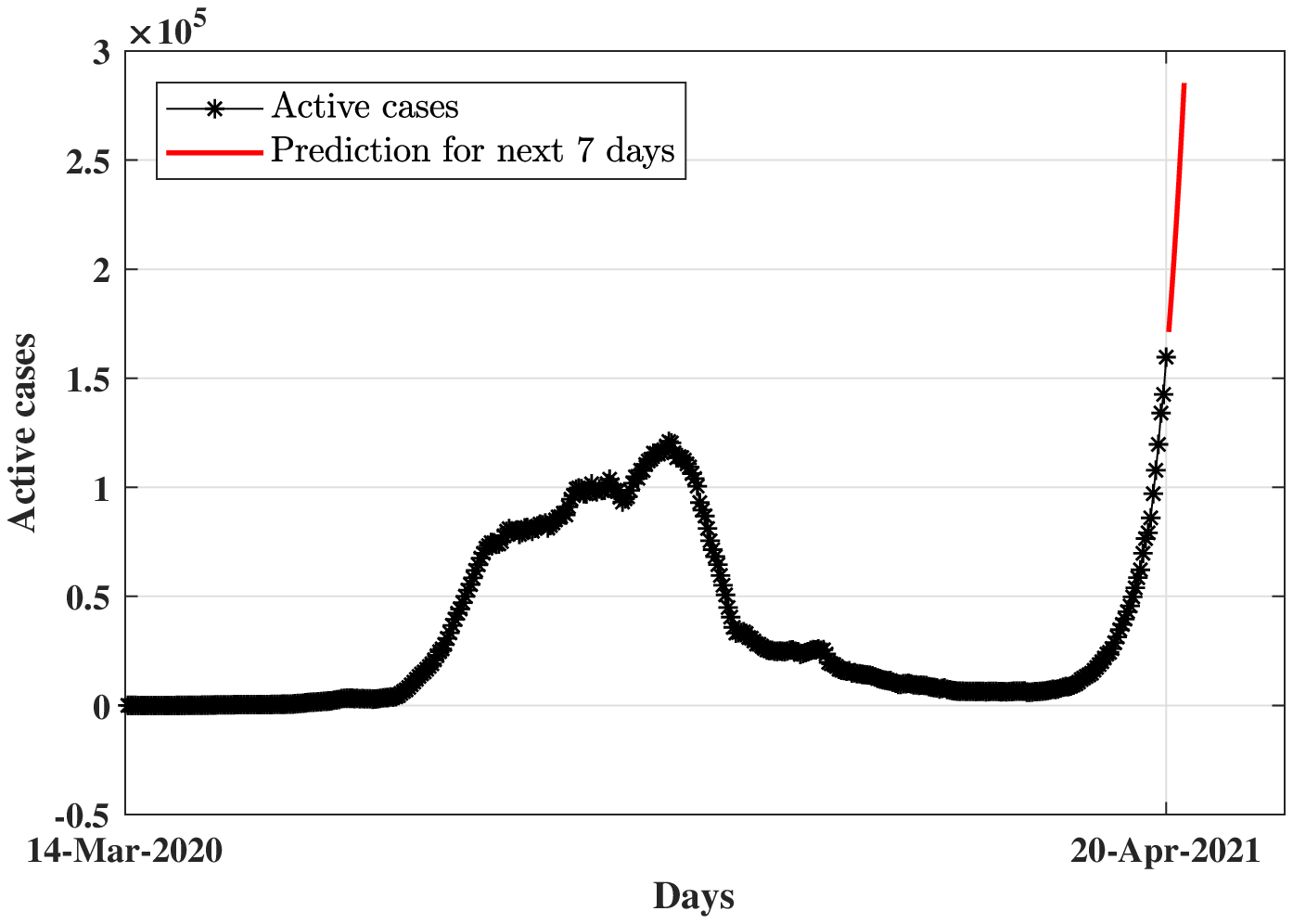}    %
\label{fig:manbeamlength1}} 
\subfigure[\textbf{Madhya Pradesh}]{%
\includegraphics[width=0.32\columnwidth, keepaspectratio]{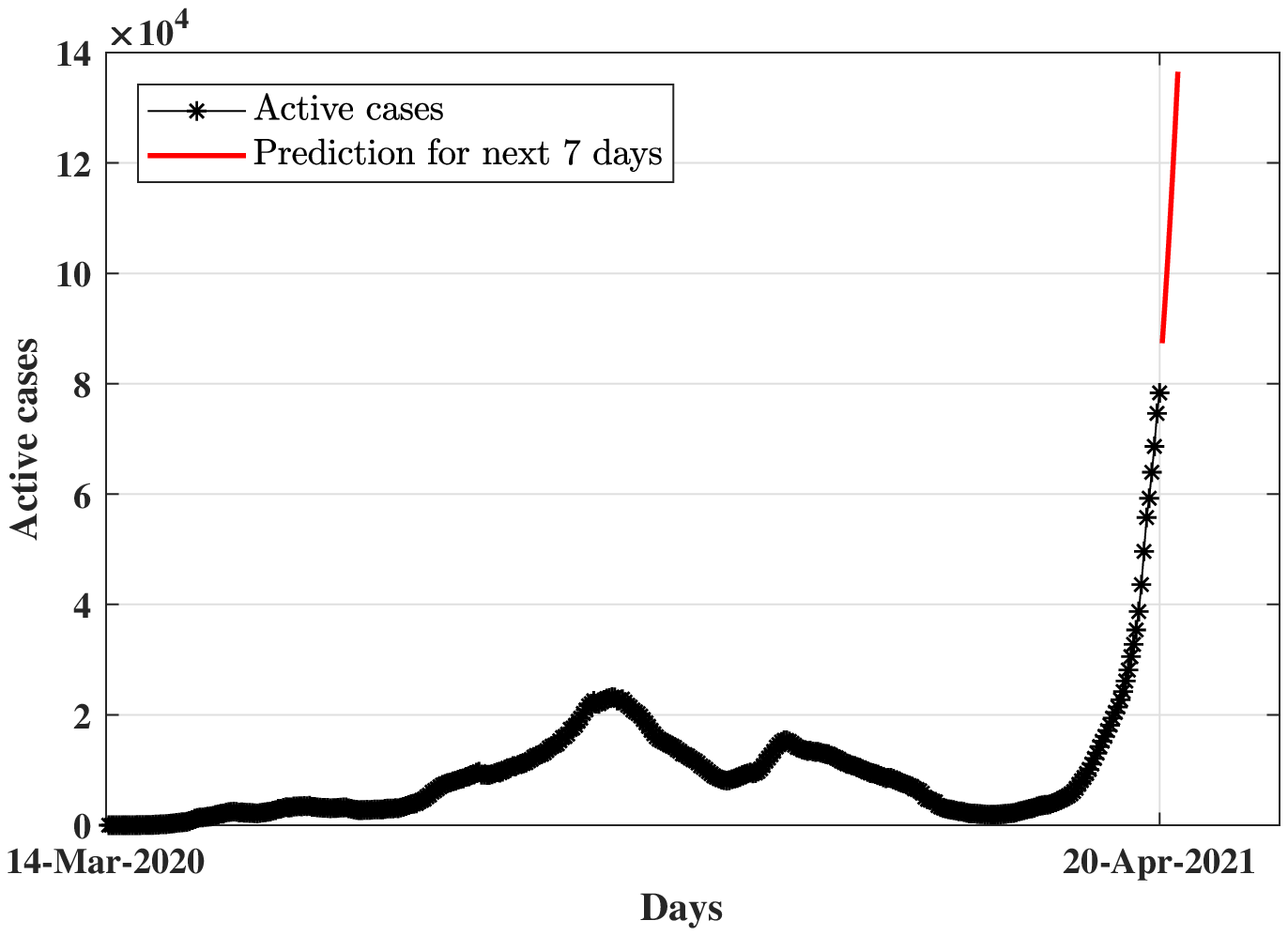}    
\label{fig:man2Dthetar1}}
\subfigure[\textbf{Delhi}]{%
\includegraphics[width=0.32\columnwidth, keepaspectratio]{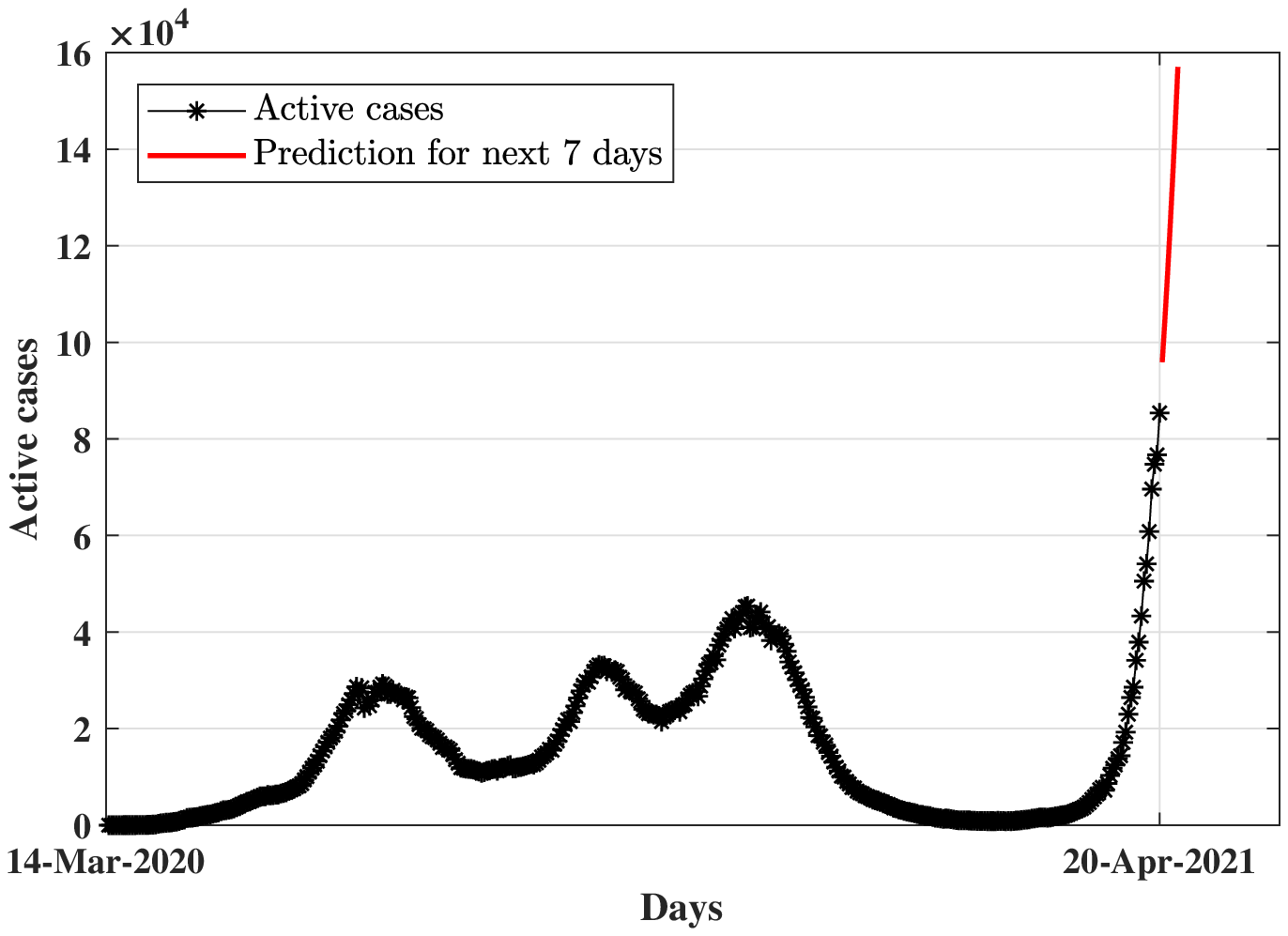}    
\label{fig:man2Dvel1}}
\subfigure[ \textbf{Haryana}]{%
\includegraphics[width=0.32\linewidth, keepaspectratio]{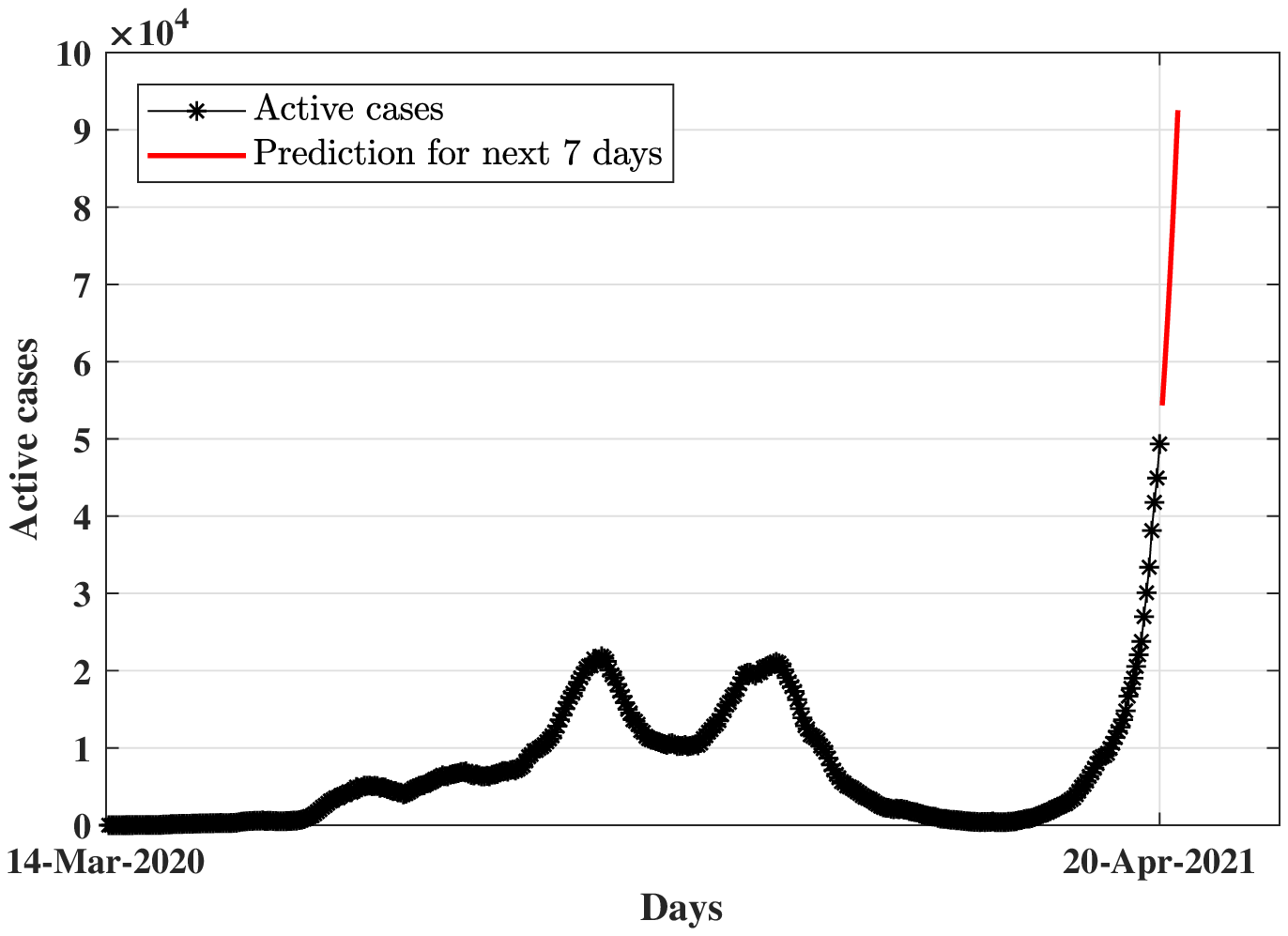}    
\label{fig:man2Dlatlatex1}}   
\subfigure[\textbf{Uttar Pradesh}]{%
\includegraphics[width=0.32\linewidth, keepaspectratio]{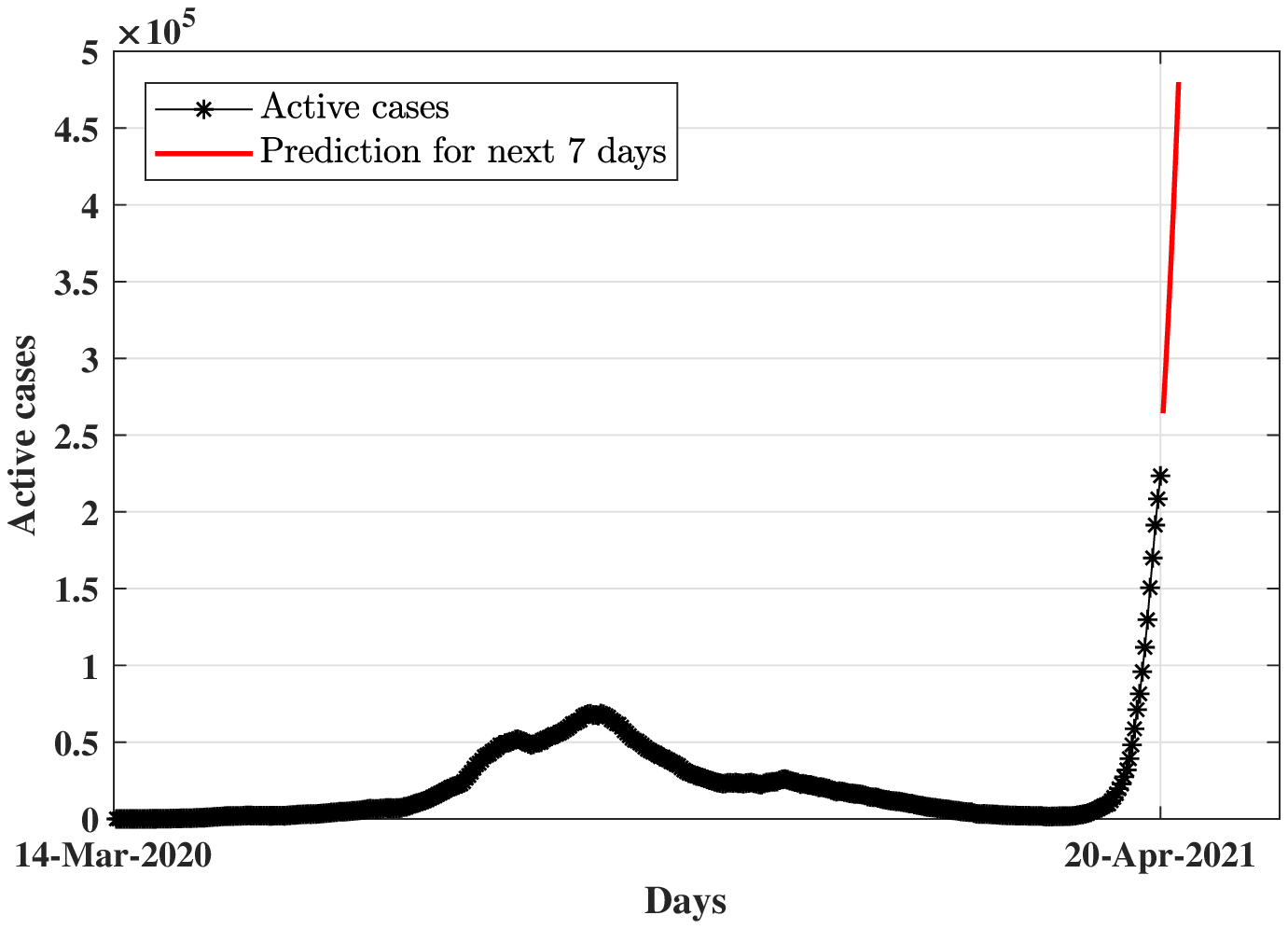}     
\label{fig:man2Dlonglatex1}}
\subfigure[\textbf{Tamil Nadu}]{%
\includegraphics[width=0.32\linewidth, keepaspectratio]{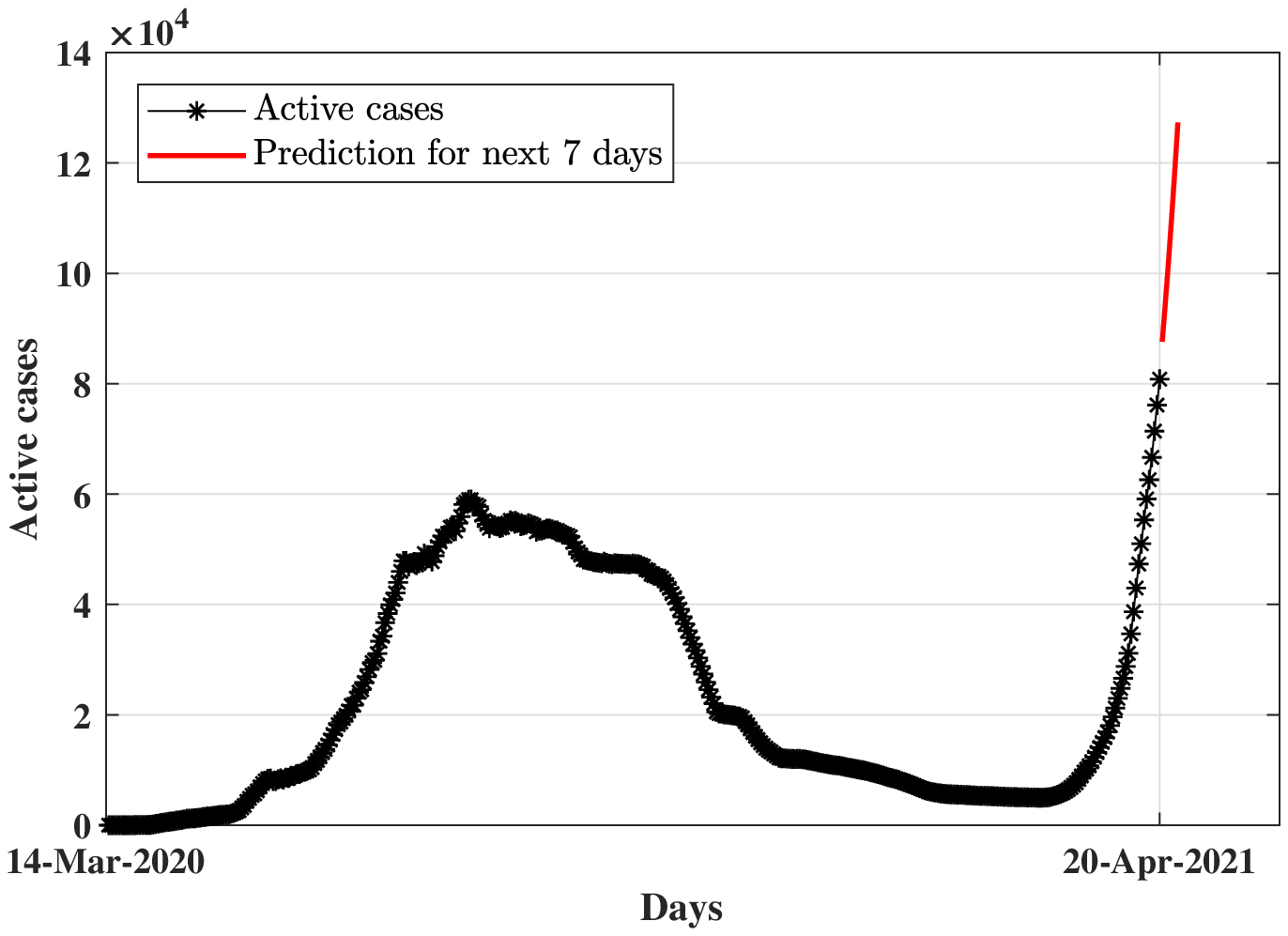}     
\label{fig:man2Dgamma1}}
\subfigure[\textbf{Kerala}]{%
\includegraphics[width=0.32\linewidth, keepaspectratio]{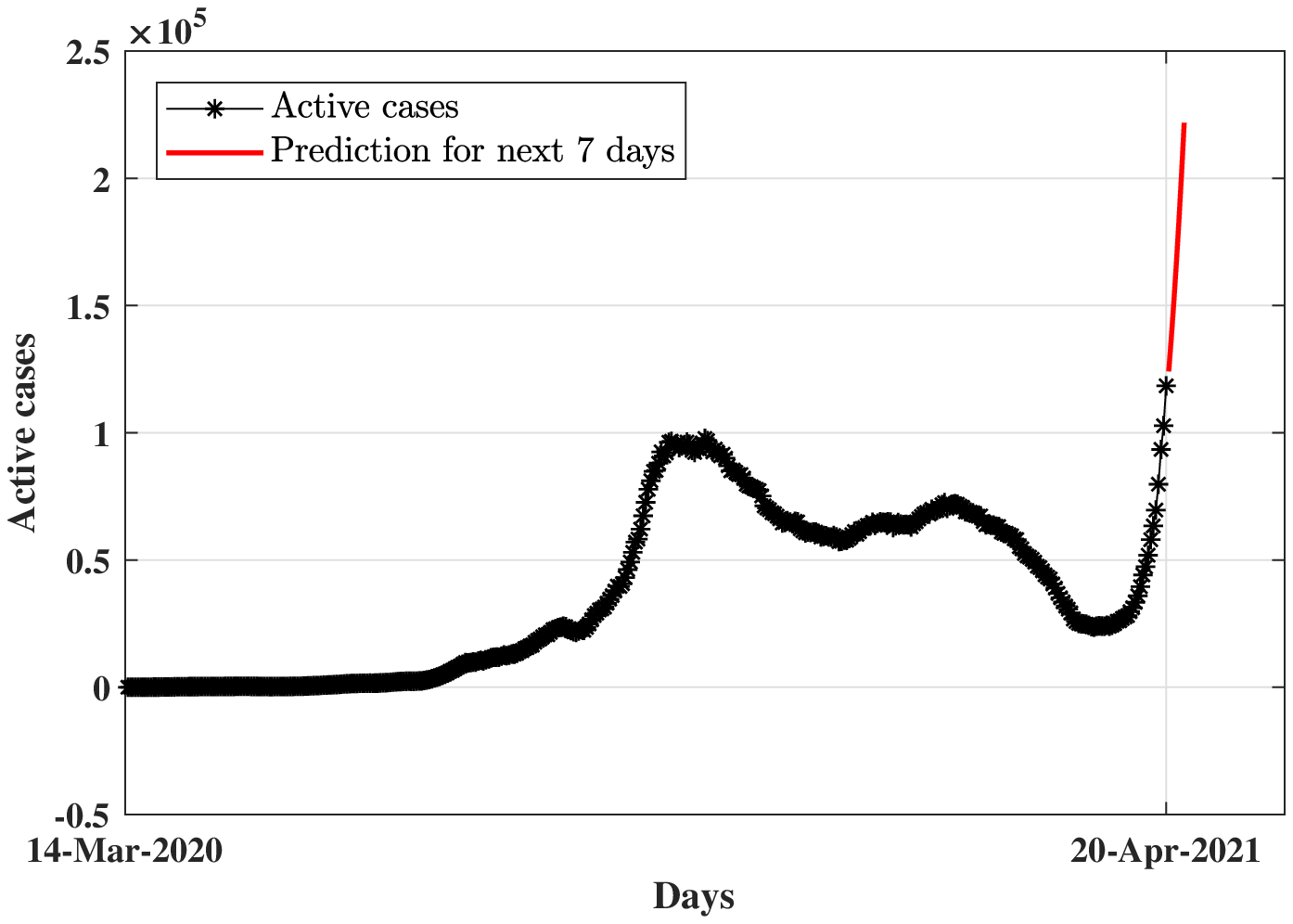}     
\label{fig:omega1}}
\caption{Predicion of active cases for next 7 days  on $20 ^{\text{th}}$ April}
\label{fig:state_1}
\end{figure}

\begin{figure}[hbt!]
\centering
\subfigure[ \textbf{Chhattisgarh}]{%
\includegraphics[width=0.32\columnwidth,keepaspectratio]{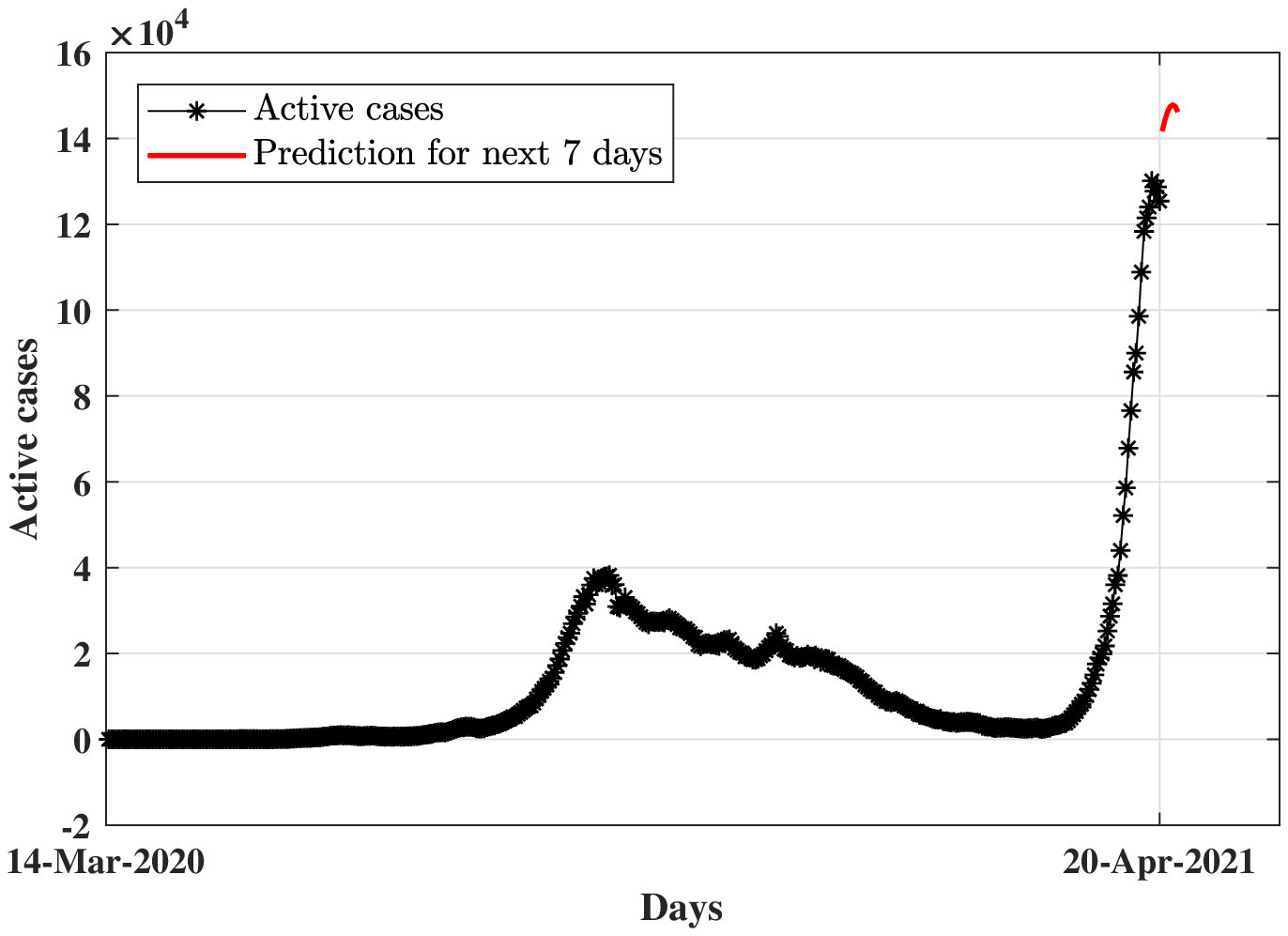}    %
\label{fig:2dtrajectorymaneuvere1}}  
\subfigure[\textbf{Rajasthan }]{%
\includegraphics[width=0.32\columnwidth, height=0.25\columnwidth]{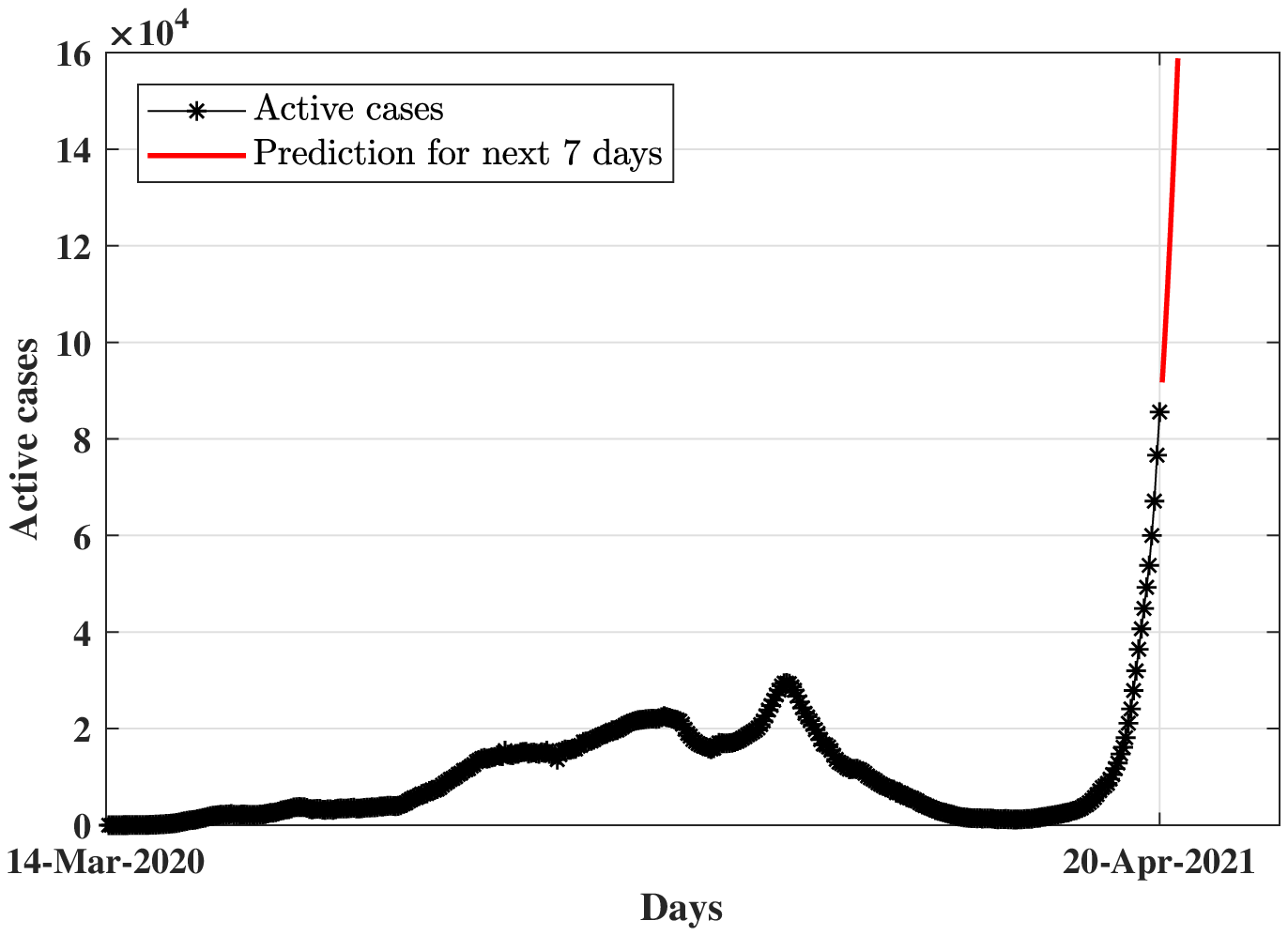}    
\label{fig:man2DR1}}
\subfigure[\textbf{ Telangana}]{%
\includegraphics[width=0.32\columnwidth, keepaspectratio]{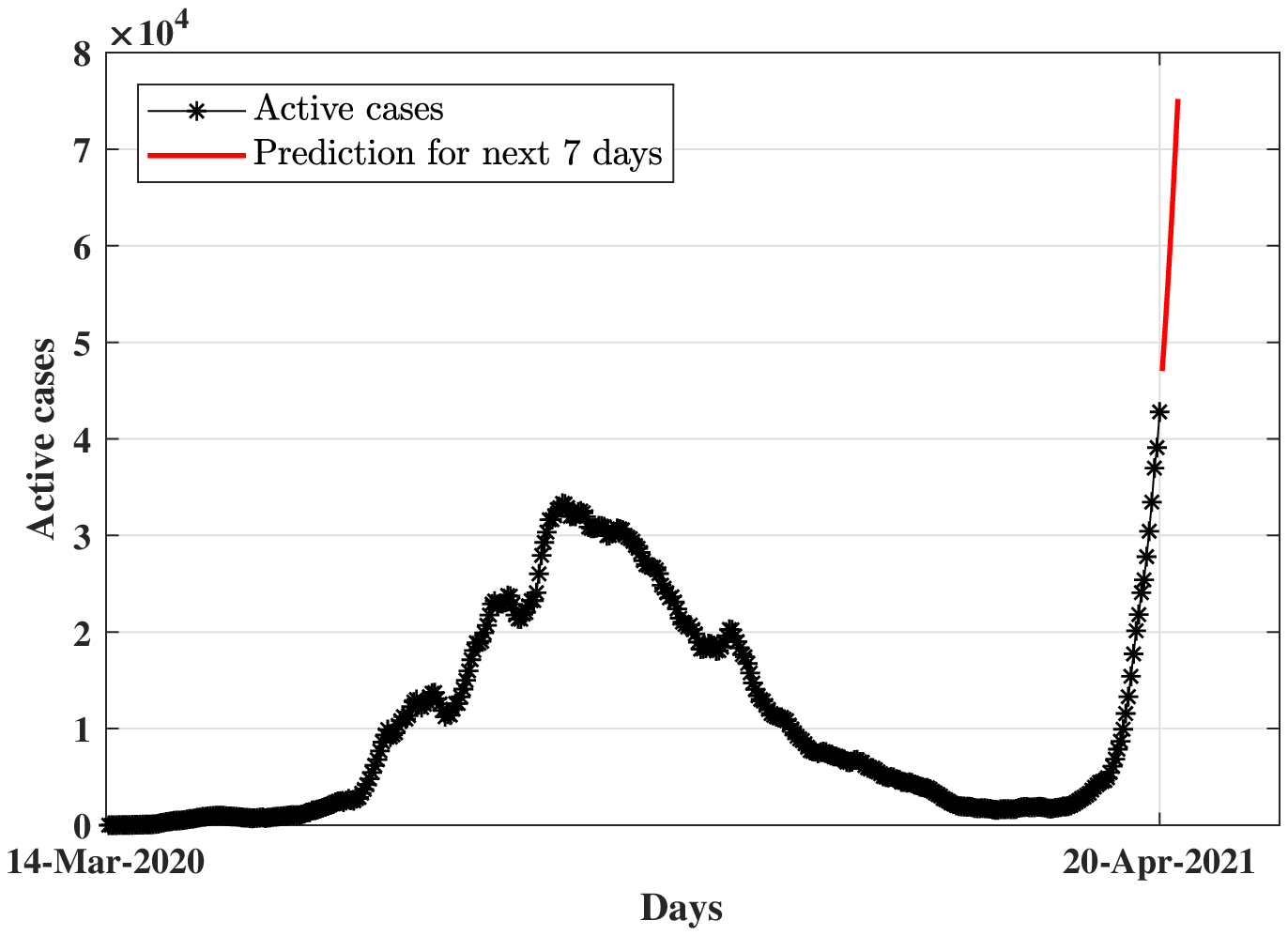}    %
\label{fig:manbeamlength1}} 
\subfigure[\textbf{Andhra Pradesh}]{%
\includegraphics[width=0.32\columnwidth, keepaspectratio]{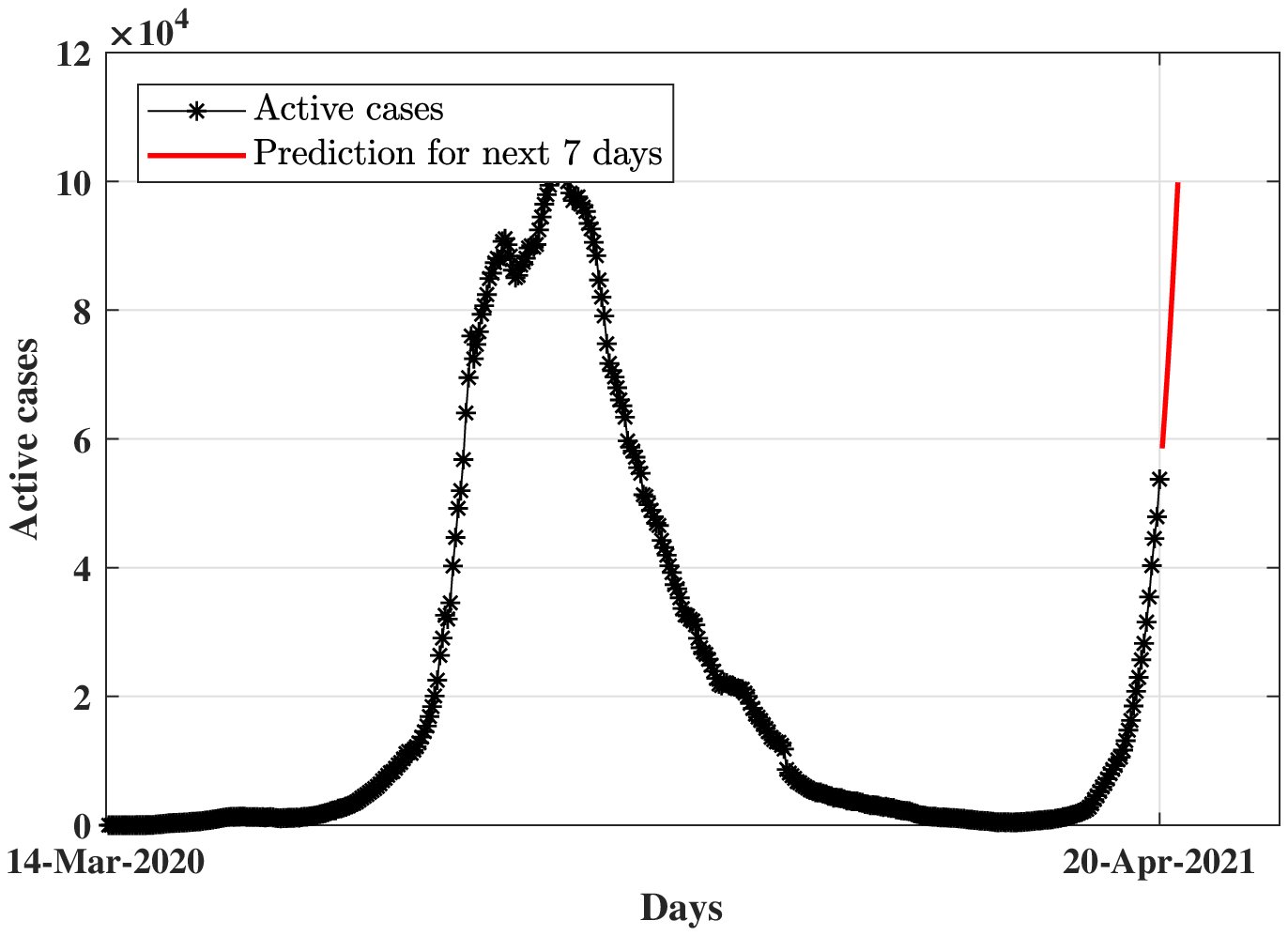}    
\label{fig:man2Dthetar1}}
\subfigure[\textbf{Uttarakhand}]{%
\includegraphics[width=0.32\columnwidth, keepaspectratio]{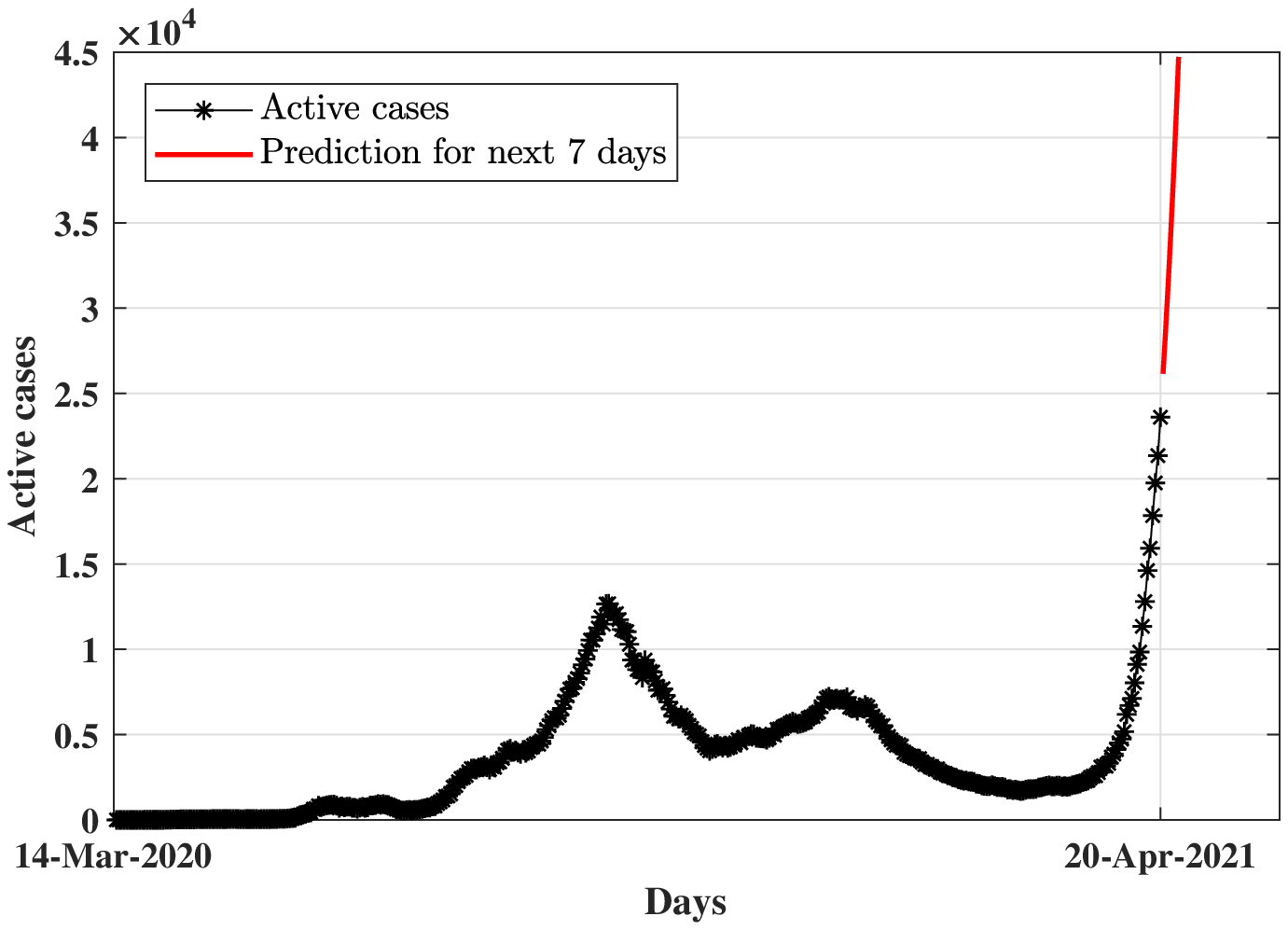}    
\label{fig:man2Dvel1}}
\subfigure[ \textbf{Jammu and Kashmir }]{%
\includegraphics[width=0.32\linewidth, keepaspectratio]{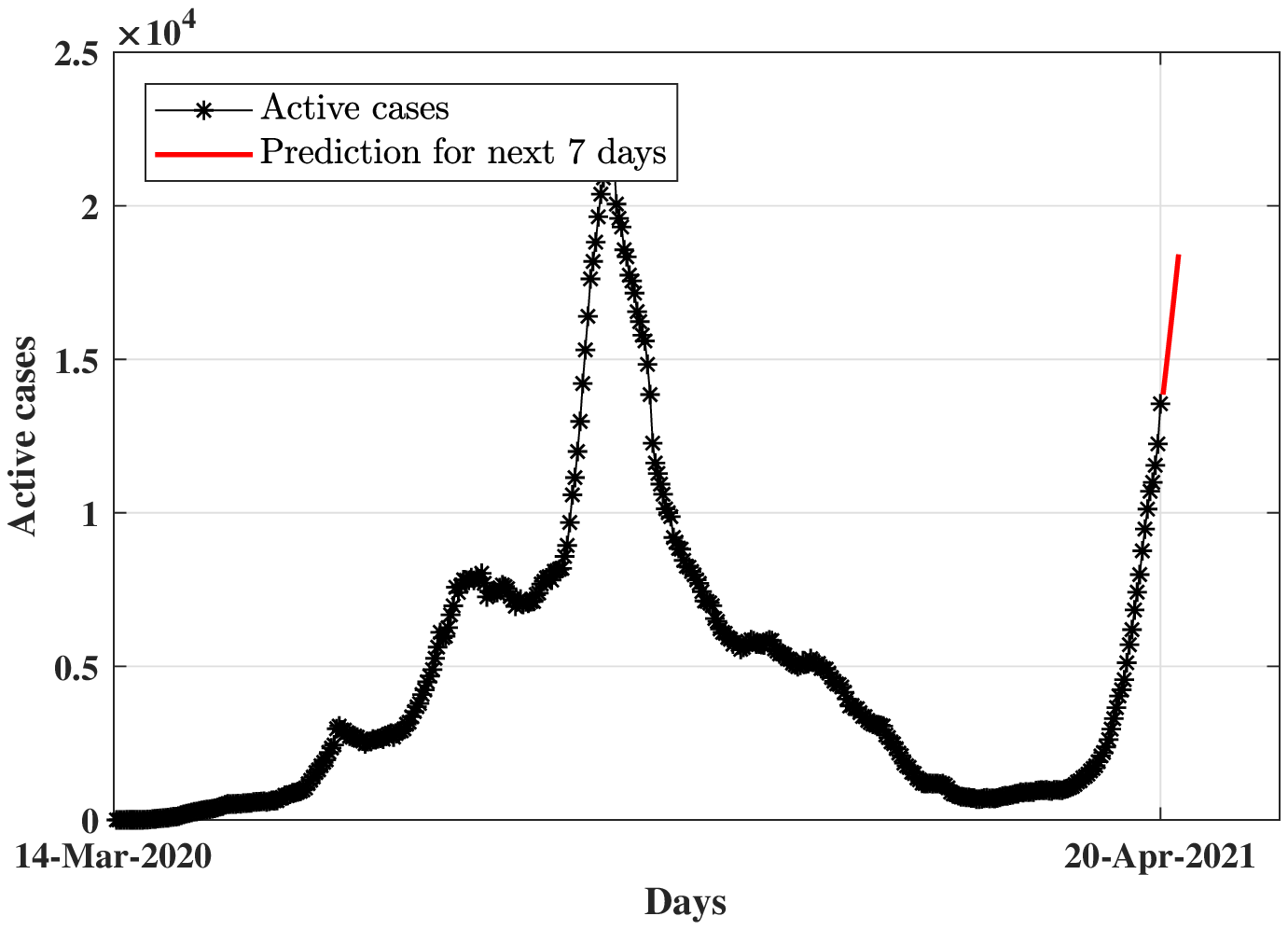}    
\label{fig:man2Dlatlatex1}}   
\subfigure[\textbf{Goa}]{%
\includegraphics[width=0.32\linewidth, keepaspectratio]{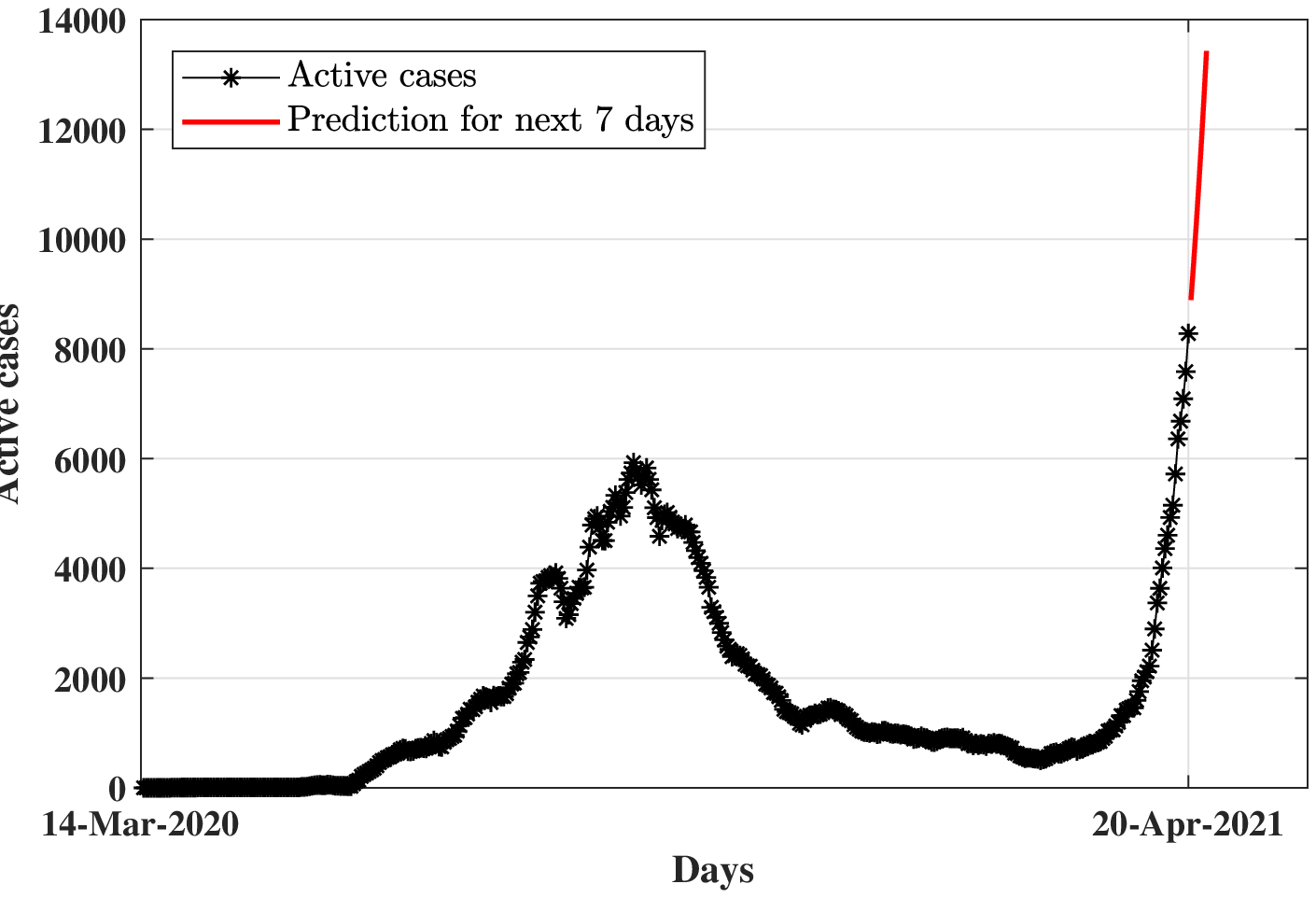}     
\label{fig:man2Dlonglatex1}}
\subfigure[\textbf{Chandigarh}]{%
\includegraphics[width=0.32\linewidth, keepaspectratio]{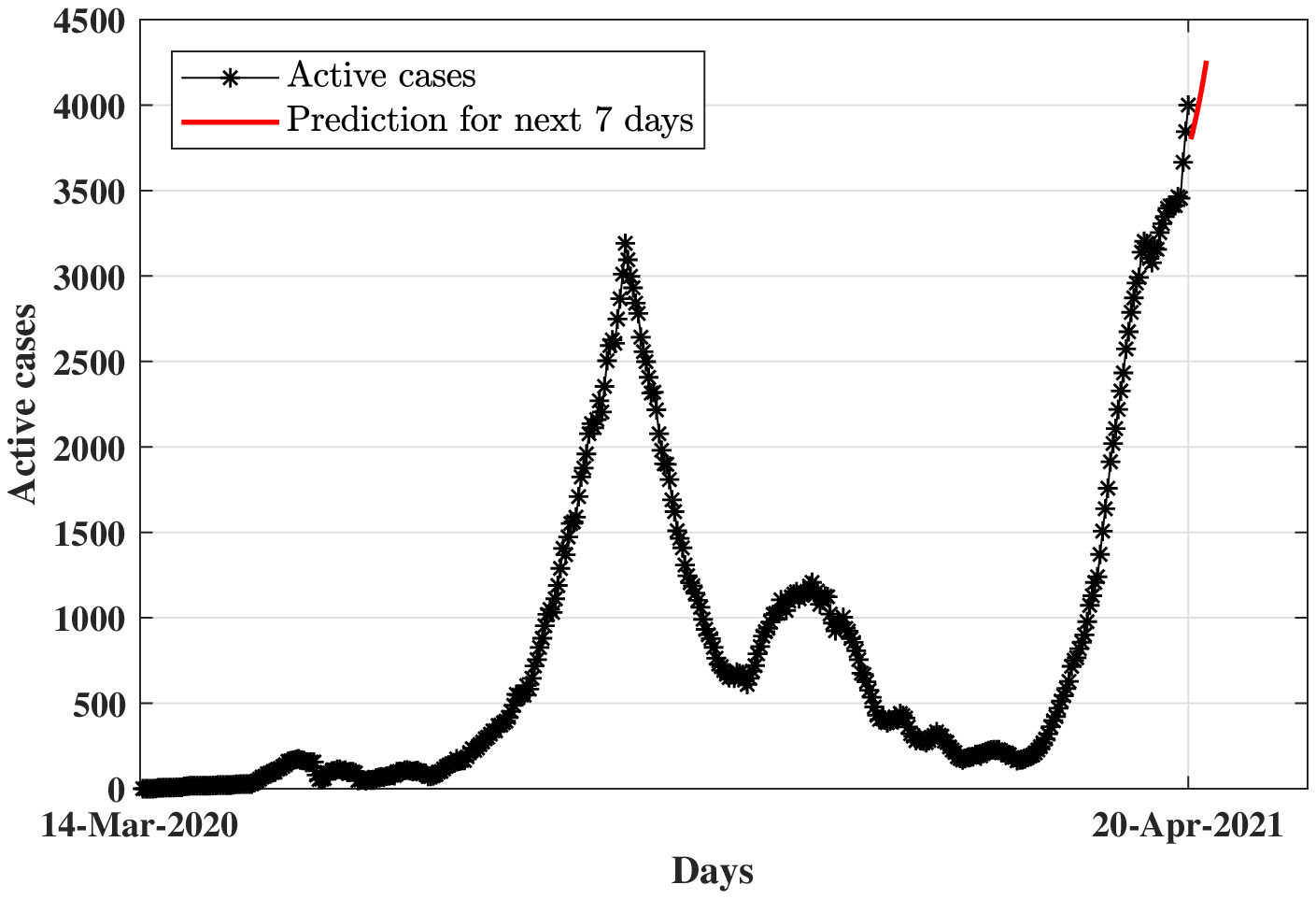}     
\label{fig:man2Dgamma1}}
\subfigure[\textbf{Himachal Pradesh}]{%
\includegraphics[width=0.32\linewidth, keepaspectratio]{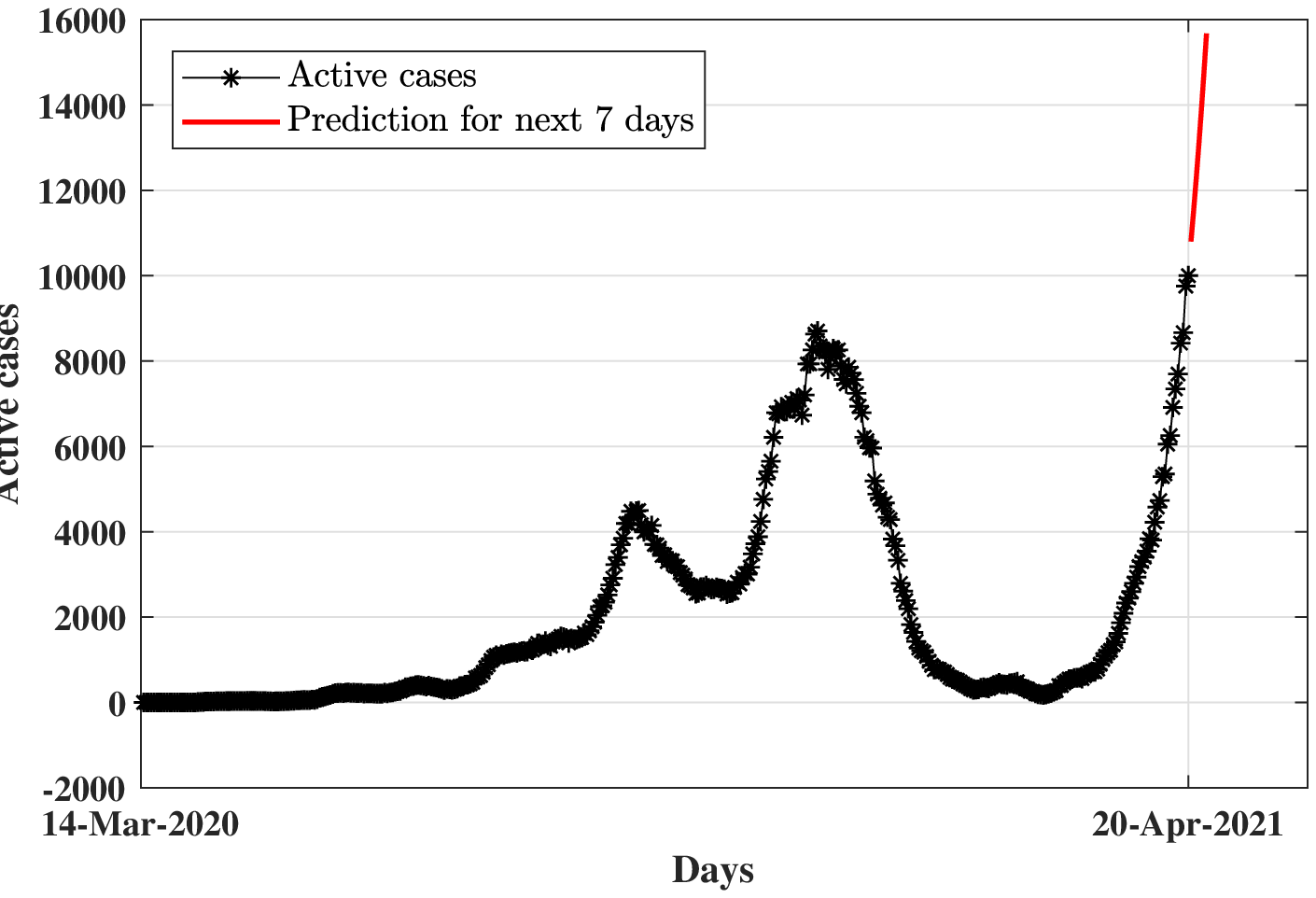}     
\label{fig:omega1}}
\caption{Predicion of active cases for next 7 days  on $20 ^{\text{th}}$ April}
\label{fig:state_2}
\end{figure}

  The maximum value of predicted active cases in a horizon of the next seven days is calculated using the time series prediction model. The prediction of active cases over  seven days is shown in Fig. \ref{fig:state_1} and Fig. \ref{fig:state_2}.  The maximum predicted value and the corresponding fraction with respect to the sum of maximum predicted values of all regions are listed in Table \ref{tab:prediction}. The corresponding allocation of the available amount of oxygen (5000 MT) is listed in the last column of Table \ref{tab:prediction}.
 
 \FloatBarrier

\begin{table}[hbt!]
\caption{\label{tab:prediction} Allocation based on maximum value of predicted active cases over next 7 days from 20 April, 2021}
\centering
\begin{tabular}{lccccc }
\hline
\hline
\bf{State} & \bf{Demand} & \bf{Active case} & \bf{Predicted value} & \bf{Weightage}  & \bf{Allocation}   \\ \hline
         Maharashtra & 1500 &  683856  & 709082  & 24.15 & 1207.5 \\
         
         Gujarat & 1000     & 76500  & 148436   &  5.06   & 253    \\ 
         
        Karnataka &  300   &  159158    & 285307 & 9.71  & 485.5 \\
        
        Madhya Pradesh &  445      & 78271   & 136516 & 4.65 &  232.5 \\

        Delhi &  700   &  85571    &   157031  & 5.35 & 267.5  \\
        
        Haryana & 180    &  49772 & 92531  &  3.15 & 157.5  \\
        
        Uttar Pradesh &  800    &  223544  & 479879 &  16.34 &  817 \\

        Tamil Nadu & 200 &  79804     &   127336 & 4.34 & 217  \\ 
        
         Kerala &  89 & 118669  & 221811  &   7.55 & 377.5 \\
        
         Chhattisgarh & 215  & 125688  & 147839 & 5.03 & 251.5 \\

         Rajasthan & 205  & 85571  & 158795 & 5.41 & 270.5\\
        
         Telangana & 360 & 42853 & 75187   & 2.56 & 128 \\ 
         
         Andhra Pradesh &  440 &  53889 &  99834 & 3.4 & 170 \\
         
         Uttarakhand & 103  & 21014  &  44727  & 1.52 &  76 \\

         Jammau and Kashmir &  12 & 13470  & 18414 & 0.63 & 31.5\\
         
          Goa & 11  & 8241    &  13428  &  0.48 & 24  \\ 
         
         Chandigarh &  20 & 3959 &  4259 & 0.14 & 7 \\
         
         Himachal Pradesh & 15   & 10029  & 15676  &  0.53 & 26.5\\

\hline
\hline
\end{tabular}
\end{table}
 
 The demand of Karnataka, Uttar Pradesh, Tamil Nadu, Kerala, Chhatisgarh, Rajasthan, Jammu and Kashmir, Goa, Himachal Pradesh is less compared to their allocation based on the weightage derived from predicted active cases over the next seven days.  Therefore, these states are allocated in full as per their demand amount.  After full allocation to these states, the available oxygen amount is 3153 MT, and the balance demand is   4748 MT.  This amount is distributed among the other states using the proposed optimization framework.  The allocated amount using the proposed framework is shown in Table \ref{tab:optimalalloc}.  The region Chandigarh and Haryana is allocated a little higher amount than their actual demand. Therefore, the excess allocation from the demand is equally distributed among the other states. The final allocation of 5000 MT   among different states is shown in  Table \ref{tab:finalalloc}.

 \begin{table}[hbt!]
\caption{\label{tab:optimalalloc}  Allocation  among states with higher demand than relative to  predicted active cases fraction}
\centering
\begin{tabular}{lccccc }
\hline
\hline
\bf{State}& Demand & \bf{Active case} & \bf{Fraction}  &  $A_{i0}$  & Allocation  \\ \hline
         Maharashtra & 1500  & 709082  & 47.94 &   1511.55 &  1326  \\
         
         Gujarat & 1000       & 148436   &  10.04   & 316.56 & 364.38    \\ 
         
         Madhya Pradesh &  445         & 136516 & 9.23 & 291.02 & 327.74 \\ 
        
         Delhi &  700     &   157031  & 10.62 & 334.85 & 388.39  \\
        
         Haryana & 180    & 92531  &  6.26 & 197.38 & 185.09  \\
        
         Telangana & 360  & 75187   &  5.08 & 160.17 & 189.8  \\ 
         
         Andhra Pradesh &  440  &  99834  & 6.75 & 212.82 & 250.08 \\
         
         Uttarakhand & 103    &  44727  & 3.02 &  95.22  & 98.03 \\
         
         Chandigarh &  20  &   4259  &  1.06  & 33.42 & 23.49\\
  
\hline
\hline
\end{tabular}
\end{table}

    \begin{table}[hbt!]
\caption{\label{tab:finalalloc}  Final oxygen allocation  of 5000 MT  for demand of 6595  MT }
\centering
\begin{tabular}{lccccc }
\hline
\hline
\bf{State}& Demand & \bf{Allocation} & \bf{Sate}  & \bf{Demand}  & Allocation  \\ \hline
         Maharashtra & 1500 &  1330.44  &  Gujarat & 1000     &  365.23  \\
         
        Karnataka &  300   &  300    & Madhya Pradesh &  445  & 328.59\\
        
        Delhi &  700   &  389.37  & Haryana & 180 & 180  \\
        
        Uttar Pradesh &  800    &  800  & Tamil Nadu & 200 &  200 \\

         Kerala &  89 & 89 &  Chhattisgarh & 215  & 215 \\

         Rajasthan & 205  & 205  &  Telangana & 360 & 190.27 \\

         Andhra Pradesh &  440 &  250.70  &  Uttarakhand & 103  & 98.30 \\

         Jammau and Kashmir &  12 & 12  & Goa & 11 & 11\\

         Chandigarh &  20 & 20  &  Himachal Pradesh & 15 &15 \\

\hline
\hline
\end{tabular}
\end{table}

 \FloatBarrier

 
 

\section{Conclusions} \label{conclusion}
This paper presents a  qualitative resource allocation framework for the distribution of critical medical resources.  The framework is developed considering the resource allocation architecture of the Indian government through the center, state,  and districts. The proposed algorithm is developed to handle the variation in unit resource requirements of different units due to different medical/administrative protocols. The framework adaptively allows allocation by accommodating the estimation of future requirements over a seven days horizon. An example is shown about allocation of oxygen among different states. The proposed architecture could be used to allocate scarce medical resources like ventilators and oxygen by any local/central authorities. 
 \FloatBarrier

\bibliographystyle{apalike}
\bibliography{main}

\begin{thebibliography}{}

\bibitem[Araz, 2013]{araz2013integrating}
Araz, O.~M. (2013).
\newblock Integrating complex system dynamics of pandemic influenza with a
  multi-criteria decision making model for evaluating public health strategies.
\newblock {\em Journal of Systems Science and Systems Engineering},
  22(3):319--339.

\bibitem[Araz et~al., 2013]{araz2013simulation}
Araz, O.~M., Lant, T., Fowler, J.~W., and Jehn, M. (2013).
\newblock Simulation modeling for pandemic decision making: A case study with
  bi-criteria analysis on school closures.
\newblock {\em Decision Support Systems}, 55(2):564--575.

\bibitem[Arora et~al., 2012]{arora2012decision}
Arora, H., Raghu, T., and Vinze, A. (2012).
\newblock Decision support for containing pandemic propagation.
\newblock {\em ACM Transactions on Management Information Systems (TMIS)},
  2(4):1--25.

\bibitem[Celik et~al., 2017]{celik2017stochastic}
Celik, E., Aydin, N., and Gumus, A.~T. (2017).
\newblock A stochastic location and allocation model for critical items to
  response large-scale emergencies: A case of turkey.
\newblock {\em An international journal of optimization and control: theories
  \& applications (IJOCTA)}, 7(1):1--15.

\bibitem[Dawson et~al., 2020]{dawson2020ethics}
Dawson, A., Isaacs, D., Jansen, M., Jordens, C., Kerridge, I., Kihlbom, U.,
  Kilham, H., Preisz, A., Sheahan, L., and Skowronski, G. (2020).
\newblock An ethics framework for making resource allocation decisions within
  clinical care: responding to {COVID}-19.
\newblock {\em Journal of bioethical inquiry}, 17(4):749--755.

\bibitem[Fogli and Guida, 2013]{fogli2013knowledge}
Fogli, D. and Guida, G. (2013).
\newblock Knowledge-centered design of decision support systems for emergency
  management.
\newblock {\em Decision Support Systems}, 55(1):336--347.

\bibitem[Govindan et~al., 2020]{govindan2020decision}
Govindan, K., Mina, H., and Alavi, B. (2020).
\newblock A decision support system for demand management in healthcare supply
  chains considering the epidemic outbreaks: A case study of coronavirus
  disease 2019 ({COVID}-19).
\newblock {\em Transportation Research Part E: Logistics and Transportation
  Review}, 138:101967.

\bibitem[G{\"u}ler and Ge{\c{c}}ici, 2020]{guler2020decision}
G{\"u}ler, M.~G. and Ge{\c{c}}ici, E. (2020).
\newblock A decision support system for scheduling the shifts of physicians
  during {COVID}-19 pandemic.
\newblock {\em Computers \& Industrial Engineering}, 150:106874.

\bibitem[Gupta and Ranganathan, 2006]{gupta2006social}
Gupta, U. and Ranganathan, N. (2006).
\newblock Social fairness in multi-emergency resource management.
\newblock In {\em IEEE International Symposium on Technology and Society
  (ISTAS), \normalfont{NY, USA}}, pages 1--9.

\bibitem[Hashemkhani~Zolfani et~al., 2020]{hashemkhani2020application}
Hashemkhani~Zolfani, S., Yazdani, M., Ebadi~Torkayesh, A., and Derakhti, A.
  (2020).
\newblock Application of a gray-based decision support framework for location
  selection of a temporary hospital during {COVID}-19 pandemic.
\newblock {\em Symmetry}, 12(6):886.

\bibitem[Herold et~al., 2021]{herold2021COVID}
Herold, D.~M., Nowicka, K., Pluta-Zaremba, A., and Kummer, S. (2021).
\newblock {COVID}-19 and the pursuit of supply chain resilience: reactions and
  “lessons learned” from logistics service providers (lsps).
\newblock {\em Supply Chain Management: An International Journal}.

\bibitem[Jana et~al., 2021]{jana2021decision}
Jana, S., Majumder, R., Menon, P.~P., and Ghose, D. (2021).
\newblock Decision support system ({DSS}) for hierarchical allocation of
  resources and tasks for disaster management.
\newblock {\em \normalfont{Presented at} \textit{5th International Conference
  on Dynamics Of Disasters (DOD)}, \normalfont{Athens, Greece}}.

\bibitem[Jenvald et~al., 2007]{jenvald2007simulation}
Jenvald, J., Morin, M., Timpka, T., and Eriksson, H. (2007).
\newblock Simulation as decision support in pandemic influenza preparedness and
  response.
\newblock {\em Proceedings ISCRAM2007}, pages 295--304.

\bibitem[Laventhal et~al., 2020]{laventhal2020ethics}
Laventhal, N., Basak, R., Dell, M.~L., Diekema, D., Elster, N., Geis, G.,
  Mercurio, M., Opel, D., Shalowitz, D., Statter, M., et~al. (2020).
\newblock The ethics of creating a resource allocation strategy during the
  {COVID}-19 pandemic.
\newblock {\em Pediatrics}, 146(1).

\bibitem[Majumder et~al., 2019]{majumder2019game}
Majumder, R., Warier, R.~R., and Ghose, D. (2019).
\newblock Game theory-based allocation of critical resources during natural
  disasters.
\newblock In {\em 2019 Sixth Indian Control Conference (ICC)}, pages 514--519.

\bibitem[Marques et~al., 2021]{marques2021prediction}
Marques, J. A.~L., Gois, F. N.~B., Xavier-Neto, J., and Fong, S.~J. (2021).
\newblock Prediction for decision support during the {COVID}-19 pandemic.
\newblock In {\em Predictive Models for Decision Support in the {COVID}-19
  Crisis}, pages 1--13. Springer.

\bibitem[Montero-Odasso et~al., 2020]{montero2020age}
Montero-Odasso, M., Hogan, D.~B., Lam, R., Madden, K., MacKnight, C., Molnar,
  F., and Rockwood, K. (2020).
\newblock Age alone is not adequate to determine health-care resource
  allocation during the {COVID}-19 pandemic.
\newblock {\em Canadian Geriatrics Journal}, 23(1):152.

\bibitem[Parsons and Johal, 2020]{parsons2020best}
Parsons, J.~A. and Johal, H.~K. (2020).
\newblock Best interests versus resource allocation: could {COVID}-19 cloud
  decision-making for the cognitively impaired?
\newblock {\em Journal of medical ethics}, 46(7):447--450.

\bibitem[Phillips-Wren et~al., 2020]{phillips2020supporting}
Phillips-Wren, G., Pomerol, J.-C., Neville, K., and Adam, F. (2020).
\newblock Supporting decision making during a pandemic: Influence of stress,
  analytics, experts, and decision aids.
\newblock {\em The Business of Pandemics: The {COVID}-19 Story}, page 183.

\bibitem[Prakash et~al., 2020]{prakash2020minimal}
Prakash, M.~K., Kaushal, S., Bhattacharya, S., Chandran, A., Kumar, A., and
  Ansumali, S. (2020).
\newblock A minimal and adaptive prediction strategy for critical resource
  planning in a pandemic.
\newblock {\em medRxiv}.

\bibitem[Ranganathan et~al., 2007]{ranganathan2007automated}
Ranganathan, N., Gupta, U., Shetty, R., and Murugavel, A. (2007).
\newblock An automated decision support system based on game theoretic
  optimization for emergency management in urban environments.
\newblock {\em Journal of Homeland Security and Emergency Management},
  4(2):1--25.

\bibitem[Sharma et~al., 2020]{sharma2020multi}
Sharma, A., Bahl, S., Bagha, A.~K., Javaid, M., Shukla, D.~K., Haleem, A.,
  et~al. (2020).
\newblock Multi-agent system applications to fight {COVID}-19 pandemic.
\newblock {\em Apollo Medicine}, 17(5):41.

\bibitem[Shearer et~al., 2020]{shearer2020infectious}
Shearer, F.~M., Moss, R., McVernon, J., Ross, J.~V., and McCaw, J.~M. (2020).
\newblock Infectious disease pandemic planning and response: Incorporating
  decision analysis.
\newblock {\em PLoS medicine}, 17(1):e1003018.

\bibitem[Sheu, 2007]{sheu2007emergency}
Sheu, J.-B. (2007).
\newblock An emergency logistics distribution approach for quick response to
  urgent relief demand in disasters.
\newblock {\em Transportation Research Part E: Logistics and Transportation
  Review}, 43(6):687--709.

\bibitem[Siraj et~al., 2020]{siraj2020infectious}
Siraj, M.~S., Dewey, R.~S., and Hassan, A. F.~U. (2020).
\newblock The infectious diseases act and resource allocation during the
  {COVID}-19 pandemic in bangladesh.
\newblock {\em Asian Bioethics Review}, 12(4):491--502.

\bibitem[Vereshchaka and Dong, 2019]{vereshchaka2019dynamic}
Vereshchaka, A. and Dong, W. (2019).
\newblock Dynamic resource allocation during natural disasters using
  multi-agent environment.
\newblock In {\em International Conference on Social Computing,
  Behavioral-Cultural Modeling and Prediction and Behavior Representation in
  Modeling and Simulation}, pages 123--132. Springer.

\bibitem[Wang et~al., 2020]{wang2020emergency}
Wang, F., Pei, Z., Dong, L., and Ma, J. (2020).
\newblock Emergency resource allocation for multi-period post-disaster using
  multi-objective cellular genetic algorithm.
\newblock {\em IEEE Access}, 8:82255--82265.

\bibitem[Worby and Chang, 2020]{worby2020face}
Worby, C.~J. and Chang, H.-H. (2020).
\newblock Face mask use in the general population and optimal resource
  allocation during the {COVID}-19 pandemic.
\newblock {\em Nature communications}, 11(1):1--9.

\end{thebibliography}

\end{document}